\documentclass[twocolumn]{aastex631}
\newcommand{\unit}[1]{\ensuremath{\, \mathrm{#1}}}
\newcommand{\gaiaFour}{Gaia-4}
\newcommand{\gaiaFive}{Gaia-5}
\newcommand{\resMgaiafour}{$M = 11.8 \pm 0.7 \unit{M_J}$} 
\newcommand{\resPgaiafour}{$P = 571.3 \pm 1.4 \unit{day}$} 
\newcommand{\resMgaiafive}{$M = 20.9 \pm 0.5\unit{M_J}$} 
\newcommand{\resPgaiafive}{$P = 358.58 \pm 0.19\unit{days}$} 
\newcommand{\resomegagaiafourb}{$\omega=180 \pm 5^{\circ}$} 
\newcommand{\resomegagaiafiveb}{$\omega=271.71 \pm 0.65^\circ$} 

\usepackage[T1]{fontenc}
\usepackage[utf8]{inputenc} 
\usepackage{comment}
\usepackage{float}

\newcommand{\PSUAA}{Department of Astronomy \& Astrophysics, 525 Davey Laboratory, The Pennsylvania State University, University Park, PA, 16802, USA}
\newcommand{\PSUARC}{Astrobiology Research Center, 525 Davey Laboratory, The Pennsylvania State University, University Park, PA, 16802, USA}
\newcommand{\PSUCEHW}{Center for Exoplanets and Habitable Worlds, 525 Davey Laboratory, The Pennsylvania State University, University Park, PA, 16802, USA}
\newcommand{\PSETI}{Penn State Extraterrestrial Intelligence Center, 525 Davey Laboratory, The Pennsylvania State University, University Park, PA, 16802, USA}
\newcommand{\UA}{Steward Observatory, The University of Arizona, 933 N.\ Cherry Ave, Tucson, AZ 85721, USA}

\newcommand{\Penn}{Department of Physics and Astronomy, University of Pennsylvania, 209 S 33rd St, Philadelphia, PA 19104, USA}

\newcommand{\GoddardESAL}{Exoplanets and Stellar Astrophysics Laboratory, NASA Goddard Space Flight Center, Greenbelt, MD 20771, USA}

\newcommand{\Goddard}{NASA Goddard Space Flight Center, 8800 Greenbelt Road, Greenbelt, MD 20771, USA}

\newcommand{\NOAO}{NSF National Optical-Infrared Astronomy Research Laboratory, 950 N.\ Cherry Ave., Tucson, AZ 85719, USA}

\newcommand{\Macquarie}{School of Mathematical and Physical Sciences, Macquarie University, Balaclava Road, North Ryde, NSW 2109, Australia}
\newcommand{\Macquariespace}{The Macquarie University Astrophysics and Space Technologies Research Centre, Macquarie University, Balaclava Road, North Ryde, NSW 2109, Australia}
\newcommand{\NIST}{National Institute of Standards \& Technology, 325 Broadway, Boulder, CO 80305, USA}
\newcommand{\CUBoulder}{Department of Physics, 390 UCB, University of Colorado, Boulder, CO 80309, USA}
\newcommand{\JPL}{Jet Propulsion Laboratory, California Institute of Technology, 4800 Oak Grove Drive, Pasadena, California 91109}

\newcommand{\UCI}{Department of Physics \& Astronomy, The University of California, Irvine, Irvine, CA 92697, USA}

\newcommand{\Princeton}{Department of Astrophysical Sciences, Princeton University, 4 Ivy Lane, Princeton, NJ 08540, USA}

\newcommand{\MCDonald}{McDonald Observatory and Department of Astronomy, The University of Texas at Austin, 2515 Speedway, Austin, TX 78712, USA}
\newcommand{\UTSpace}{Center for Planetary Systems Habitability, The University of Texas at Austin, 2515 Speedway, Austin, TX 78712, USA}
\newcommand{\ASTRON}{ASTRON, Netherlands Institute for Radio Astronomy, Oude Hoogeveensedijk 4, Dwingeloo, 7991 PD, The Netherlands}
\newcommand{\API}{Anton Pannekoek Institute for Astronomy, University of Amsterdam, Science Park 904, 1098 XH Amsterdam, The Netherlands}
\newcommand{\Aarhus}{Stellar Astrophysics Centre, Department of Physics and Astronomy, Aarhus University, Ny Munkegade 120, 8000 Aarhus C, Denmark}

\newcommand{\TIFR}{Department of Astronomy and Astrophysics, Tata Institute of Fundamental Research, Homi Bhabha Road, Colaba, Mumbai 400005, India}
\newcommand{\LeidenObservatory}{Leiden Observatory, Leiden University, PO Box 9513, 2300 RA, Leiden, The Netherlands}
\newcommand{\PSUICDS}{Institute for Computational and Data Sciences, The Pennsylvania State University, University Park, PA, 16802, USA}
\newcommand{\FlatironCCA}{Center for Computational Astrophysics, Flatiron Institute, 162 Fifth Avenue, New York, NY 10010, USA}
\newcommand{\UIUC}{Department of Astronomy, University of Illinois at Urbana-Champaign, Urbana, IL 61801, USA}

\begin{document}

\title{Gaia-4b and 5b: Radial Velocity Confirmation of Gaia Astrometric Orbital Solutions Reveal a Massive Planet and a Brown Dwarf Orbiting Low-mass Stars}
\author[0000-0001-7409-5688]{Guðmundur Stefánsson}
\affil{\API} 

\author[0000-0001-9596-7983]{Suvrath Mahadevan}
\affil{\PSUAA}
\affil{\PSUARC}
\affil{\PSUCEHW}

\author[0000-0002-4265-047X]{Joshua N.\ Winn}
\affil{\Princeton}

\author[0000-0003-2173-0689]{Marcus L. Marcussen}
\affil{\Aarhus}

\author[0000-0001-8401-4300]{Shubham Kanodia}
\affil{Earth and Planets Laboratory, Carnegie Institution for Science, 5241 Broad Branch Road, NW, Washington, DC 20015, USA}

\author[0000-0003-1762-8235]{Simon Albrecht}
\affil{\Aarhus}

\author[0000-0003-0199-9699]{Evan Fitzmaurice}
\affiliation{\PSUAA}
\affiliation{\PSUCEHW}
\affiliation{\PSUICDS}
\affiliation{Institute for Computational and Data Sciences Scholar}

\author{Onė Mikulskitye}
\affiliation{\API}

\author[0000-0003-4835-0619]{Caleb I. Ca\~nas}
\affil{NASA Postdoctoral Fellow}
\affil{\Goddard}

\author[0000-0001-9480-8526]{Juan I. Espinoza-Retamal}
\affiliation{Instituto de Astrof\'isica, Pontificia Universidad Cat\'olica de Chile, Av. Vicu\~na Mackenna 4860, 782-0436 Macul, Santiago, Chile}
\affiliation{Millennium Institute for Astrophysics, Santiago, Chile}
\affil{\API}

\author[0009-0000-3972-0775]{Yiri Zwart}
\affiliation{\API}

\author[0000-0001-9626-0613]{Daniel M. Krolikowski}
\affiliation{\UA}

\author[0009-0000-1825-4306]{Andrew Hotnisky}
\affil{\PSUAA}
\affil{\PSUCEHW}

\author[0000-0003-0149-9678]{Paul Robertson}
\affil{\UCI}

\author[0000-0003-0353-9741]{Jaime A. Alvarado-Montes}
\affil{\Macquarie}
\affil{\Macquariespace}

\author[0000-0003-4384-7220]{Chad F.\ Bender}
\affil{\UA}

\author[0000-0002-6096-1749]{Cullen H.\ Blake}
\affil{\Penn}

\author[0000-0002-7167-1819]{J. R. Callingham}
\affil{\LeidenObservatory}
\affil{\ASTRON}

\author[0000-0001-9662-3496]{William D. Cochran}
\affil{\MCDonald}
\affil{\UTSpace}

\author[0000-0003-1439-2781]{Megan Delamer}
\affil{\PSUAA}
\affil{\PSUCEHW}

\author[0000-0002-2144-0764]{Scott A.\ Diddams}
\affil{\NIST}
\affil{\CUBoulder}

\author[0000-0002-3610-6953]{Jiayin Dong}
\affil{\FlatironCCA}
\affil{\UIUC}

\author[0000-0002-3853-7327]{Rachel B. Fernandes}
\altaffiliation{President's Postdoctoral Fellow}
\affil{\PSUAA}
\affil{\PSUCEHW}

\author[0000-0002-0078-5288]{Mark R. Giovinazzi}
\affil{Department of Physics and Astronomy, Amherst College, Amherst, MA 01002, USA}

\author[0000-0003-1312-9391]{Samuel Halverson}
\affil{\JPL}

\author[0000-0002-2990-7613]{Jessica Libby-Roberts}
\affil{\PSUAA}
\affil{\PSUCEHW}

\author[0000-0002-9632-9382]{Sarah E.\ Logsdon}
\affil{\NOAO}

\author[0000-0003-0241-8956]{Michael W.\ McElwain}
\affil{\GoddardESAL} 

\author[0000-0001-8720-5612]{Joe P.\ Ninan}
\affil{\TIFR}

\author[0000-0002-2488-7123]{Jayadev Rajagopal}
\affil{\NOAO}

\author[0009-0006-7298-619X]{Varghese Reji}
\affil{\TIFR}

\author[0000-0001-8127-5775]{Arpita Roy}
\affil{Astrophysics \& Space Institute, Schmidt Sciences, New York, NY 10011, USA}

\author[0000-0002-4046-987X]{Christian Schwab}
\affil{\Macquarie}

\author[0000-0001-6160-5888]{Jason T.\ Wright}
\affil{\PSUAA}
\affil{\PSUCEHW}
\affil{\PSETI}

\begin{abstract} 
Gaia astrometry of nearby stars is precise enough to detect the tiny displacements induced by substellar companions, but radial velocity data are needed for definitive confirmation. Here we present radial velocity follow-up observations of 28 M and K stars with candidate astrometric substellar companions, which led to the confirmation of two systems, Gaia-4b and Gaia-5b, and the refutation of 21 systems as stellar binaries. Gaia-4b is a massive planet ($M = 11.8 \pm 0.7 \:\mathrm{M_J}$) in a $P = 571.3 \pm 1.4\:\mathrm{day}$ orbit with a projected semi-major axis $a_0=0.312 \pm 0.040\:\mathrm{mas}$ orbiting a $0.644 \pm 0.02 \:\mathrm{M_\odot}$ star. Gaia-5b is a brown dwarf ($M = 20.9 \pm 0.5\:\mathrm{M_J}$) in a $P = 358.58 \pm 0.19\:\mathrm{days}$ eccentric $e=0.6412 \pm 0.0027$ orbit with a projected angular semi-major axis of $a_0 = 0.947 \pm 0.038\:\mathrm{mas}$ around a $0.34 \pm 0.03 \mathrm{M_\odot}$ star. Gaia-4b is one of the first exoplanets discovered via the astrometric technique, and is one of the most massive planets known to orbit a low-mass star.
\end{abstract}


\section{Introduction} \label{sec:intro}
\noindent The astrometric and radial velocity (RV) methods for detecting exoplanets are complementary because they measure different aspects of a star's motion around the center of mass of a planetary system. Astrometry measures the star's angular position on the sky plane, while RVs measure the line-of-sight component of its velocity. At this point, the RV method is more ubiquitous and has resulted in hundreds of exoplanet discoveries, far more than the astrometric method \citep{curiel2022}. However, the astrometric method has the virtue of being able to determine almost all the planet's orbital parameters including its mass, whereas the RV method cannot determine the orbital inclination and only reveals the planet's minimum possible mass. Furthermore, the astrometric method is poised for a major leap forward because of the Gaia mission \citep{gaia2016}. Gaia astrometric data should enable the detection of thousands of exoplanets \citep{perryman2014,espinoza2023}, many of which will be giant planets orbiting nearby low-mass stars \citep{sozzetti2014}. The resulting sample of giant planets at intermediate orbital distances could yield important clues to the formation of such planets around M and K stars which are known to be intrinsically rare \citep[e.g.,][]{endl2006,sabotta2021,pinamonti_hades_2022,gan2023,bryant2023}.

Gaia Data Release 3 \citep[DR3;][]{gaia2023vallenari} was the first data release including a `non-single star' (NSS) catalog consisting of unresolved sources for which the observed astrometric motion is consistent with a Keplerian orbit. A total of 169,227 astrometric NSS solutions were reported \citep{halbwachs2023}. The parameters of the astrometric orbit can be used to calculate the implied mass of the companion, assuming the source is a single star and the companion is dark. Under those assumptions, about 1\% of the NSS solutions imply a substellar companion mass. The Gaia team scrutinized these systems and designated 72 as `Gaia AStrometric Objects of Interest' (ASOI) for being especially promising candidates for substellar companions \citep{holl2023,arenou2023}. After a recent announcement of a software bug, the astrometric solutions for 3 of the 72 ASOIs were retracted\footnote{See \url{https://www.cosmos.esa.int/web/gaia/dr3-known-issues}.}.

However, from the Gaia-ASOI astrometric solutions alone, it is unclear if the assumptions of a single star and a single dark companion are valid. An unresolved source with a small observed photocenter amplitude might also be caused by nearly equal-mass binary star, in which the astrometric motion can mimic the orbit caused by a substellar companion. Blending of light from physically unrelated sources can also diminish the observed amplitude of astrometric motion and mimic the signal of a substellar companion \citep{marcussen2023}. Furthermore, the presence of more than one companion would cause the tabulated parameters in the NSS catalog to be biased, since the parameters were derived by fitting a single Keplerian orbit to the astrometric data.

To investigate these issues for the Gaia-ASOI systems, \cite{holl2023} and \cite{winn2022} discussed a few cases for which RVs were available in the literature. In some cases, the RVs and Gaia solutions showed good agreement, while others showed inconsistencies. One of the puzzling systems was HIP 66074b (Gaia-3b), which has a claim to being the first astrometrically detected exoplanet, but the situation is complicated. Both the Gaia and RV data were consistent with a planet in a 300 day eccentric orbit, but the observed amplitude of the RV signal was 15 times smaller than predicted from the Gaia solution \citep{winn2022,marcussen2023}. Follow-up RV observations by \cite{sozzetti2023} showed that the companion is likely substellar but the orbit is more eccentric than implied by Gaia solution ($e=0.95$ vs.\ 0.5). This system proved to be one of those for which the aforementioned software bug affected the quality of the Gaia astrometric solution leading to its withdrawal. Other efforts to characterize Gaia NSS systems have included follow-up observations of brown dwarf candidates by \cite{unger2023} and \cite{fitzmaurice2023} with the RV technique, and by \cite{winterhalder2024} with GRAVITY interferometric imaging.

Here we announce the discovery of two substellar companions to nearby low-mass stars, based on the concordance between Gaia astrometric solutions and follow-up RV observations. Gaia-4b is an early M or late K dwarf hosting a \resMgaiafour{} mass planet. Given the situation with HIP\,66074b described above, Gaia-4b appears to be the first exoplanet discovered on the basis of a valid astrometric solution. Gaia-5b is a brown dwarf orbiting a nearby mid M dwarf on an eccentric $e=0.64$ orbit. These two systems emerged from our survey of 28 M and K stars with candidate astrometric substellar companions, most of which (21) turned out to be spectroscopic double-lined binaries.

This paper is structured as follows. We discuss the observations in Section \ref{sec:obs}, and the parameters of the Gaia-4b and Gaia-5b host stars in Section \ref{sec:stellar}. We describe our modeling procedure in Section \ref{sec:modeling}, and the results in Section \ref{sec:results}. We compare Gaia-4b and Gaia-5b with other known substellar companions of M and K stars in Section \ref{sec:discussion}, and we also discuss the overall false positive rate of the Gaia ASOI candidates. We conclude with a summary of our findings in Section \ref{sec:conclusion}.

\section{Observations} \label{sec:obs}

\subsection{Gaia Astrometry}
Since its launch in 2013 Gaia has been observing the entire sky as dictated by its `scanning law' \citep{gaia2016}. As stars move across the focal plane of each of the two Gaia telescopes, their one-dimensional coordinates are measured with high precision, and the results are assembled into a global astrometric model. The DR3 results were based on 34 months of observations and about 40 scans per star, on average. As noted earlier, for a subset of the stars, the observed astrometric motion consisted not only of the usual parallax and proper motion effects but also two-body Keplerian orbital motion \citep{holl2023,arenou2023}. We retrieved the Gaia two-body solutions for all of the Gaia ASOI candidates from the Gaia archive\footnote{\url{https://gea.esac.esa.int/archive/}} and obtained the covariance matrix for the parameters using the \texttt{nsstools}\footnote{\url{https://www.cosmos.esa.int/web/gaia/dr3-nss-tools}} code.

\subsection{Spectroscopic Observations}
\begin{figure*}[t!]
\begin{center}
\includegraphics[width=0.75\textwidth]{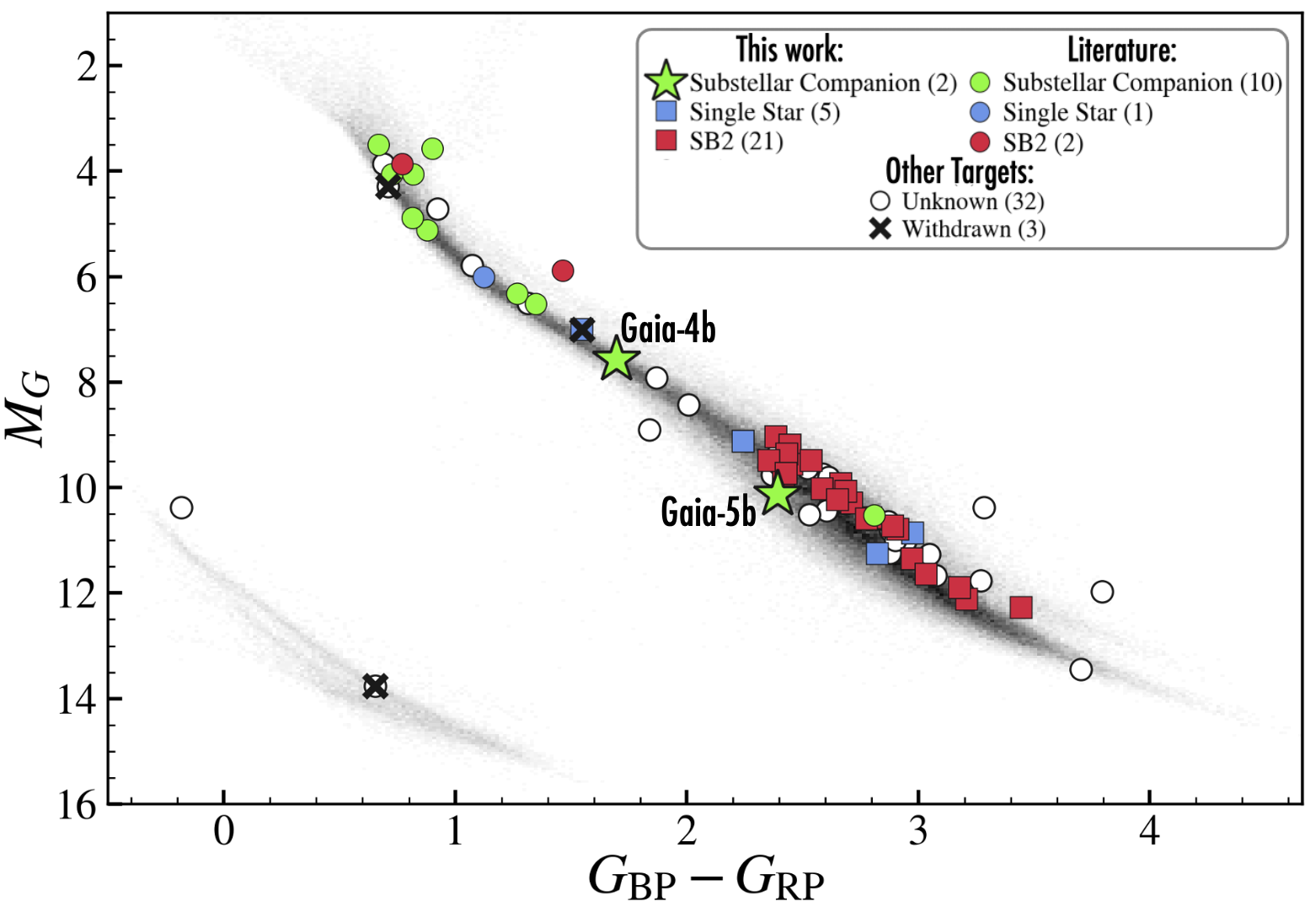}
\end{center}
\vspace{-0.5cm}
\caption{Color-magnitude diagram of Gaia astrometric candidate planets and substellar objects. The 28 systems observed as part of this work are highlighted with stars or squares, while other systems from the literature are highlighted with the circles. Confirmed substellar companions are highlighted in green. Systems that are consistent with single stars but have yet to have a Gaia astrometric companion fully confirmed are highlighted in blue. Double-lined binary systems (SB2) are highlighted in red. Systems that have an unknown designation are shown with white circles. The three Gaia-ASOI systems for which the Gaia astrometric solution was retracted are shown with black crosses. Gaia-4b and Gaia-5b are highlighted with the green star markers. For comparison, the Gaia Catalog of Nearby Stars \citep{smart2021} is shown with faint gray points in the background.}
\label{fig:cmd}
\end{figure*}

To detect potentially massive substellar companions around nearby M and K stars being unveiled by Gaia, we conducted spectroscopic observations of 28 northern-hemisphere Gaia ASOI candidates to characterize them and rule out false positives. These 28 systems constitute the northern M and K stars in the Gaia-ASOI list with a declination $\delta \gtrsim -20^\circ$ and $G_{\mathrm{Bp}} - G_{\mathrm{Rp}} > 1.5$. Figure \ref{fig:cmd} gives an overview of the current Gaia ASOI list in a color-magnitude diagram. The 28 Gaia ASOIs observed as part of this work are highlighted with square markers, and Gaia-4b and Gaia-5b are shown with star markers. We highlight sources that are observed to be double-lined spectroscopic binaries (SB2) in red, systems that are still consistent with being single stars in blue, and systems which were previously known to host planets based on RV discoveries in green circles. We discuss the false positives and the designation in further detail in Section \ref{sec:discussion} and in Table \ref{tab:targets} in the Appendix.

\subsubsection{HPF}
The Habitable-zone Planet Finder (HPF) is a temperature stabilized \citep{stefansson2016} fiber fed \citep{kanodia2018fiber} near-infrared spectrograph on the 10m Hobby-Eberly Telescope \citep[HET;][]{ramsey1998,hill2021} at McDonald Observatory in Texas covering the wavelength range from 810-1280nm at a resolution of $R\sim 55,000$ \citep{mahadevan2012,mahadevan2014}.  We obtained HPF spectra of 21 Gaia ASOI candidates around M and K dwarfs, with the goal of verifying the assumption that the Gaia source is a single star with a single dark companion. The exposure time per target was generally 900s, which was sufficient to get enough SNR to distinguish between double and single lines in a system. All of the observations were obtained as part of the HET queue \citep{shetrone2007}. The HPF 1D spectra were reduced using the HPF pipeline following procedures described by \cite{ninan2019}, \cite{kaplan2018}, and \cite{metcalf2019}. For Gaia-4 and Gaia-5, we extracted precise radial velocities from the 1D spectra using the SpEctrum Radial Velocity AnaLyzer (SERVAL) code \citep{zechmeister2018} as adapted for HPF \citep{stefansson2020,stefansson2023}. The RVs are further discussed in Section \ref{sec:results}.

To look for evidence of double-lined binaries, we calculated spectral-line broadening functions \citep{rucinski1992} for all of the HPF spectra. The broadening function is the convolution kernel that, when applied to a template spectrum, best reproduces the observed spectrum. For a single star, the broadening function has a single peak. For a double star, there will generally be two peaks unless the two radial velocities happen to be too close to be resolved. We calculated the broadening functions by adapting the \texttt{saphires}\footnote{\url{https://github.com/tofflemire/saphires}} code \citep{tofflemire2019} for use for HPF data. We briefly discuss the calculation of the HPF broadening functions here, and refer to \cite{tofflemire2019} for further details on the algorithm, which performs the convolution inversion using singular-value decomposition following \cite{rucinski1999}. For the template spectrum, we used a high-SNR spectrum of the M dwarf GJ 436, a slow rotator, which was selected as it was close in spectral type to the observed stars. By using a real spectrum instead of a theoretical model spectrum we bypassed the need for a model of instrumental broadening. We note that we experimented with using different template stars, none of which changed any of our conclusions---the main goal of the broadening functions was to look for evidence of single or double lines in the resulting broadening functions to distinguish between single stars and binary stars. For each target, we calculated the broadening function using HPF order index 5 (866--876\,nm) because of the low level of telluric absorption and sky emission lines, obviating the need for telluric corrections. We experimented with additional orders, but this single clean order proved sufficient to distinguish between single or double-lined systems. The results from the broadening function calculations are further discussed in Section \ref{sec:discussion}.

For Gaia-4 and Gaia-5, we obtained 11 and 17 observations with an average SNR of 97 and 84 per extracted 1D pixel at $1\unit{\mu m}$, respectively. This resulted in median RV uncertainties of $16 \mathrm{m\:s^{-1}}$ and $14 \mathrm{m\:s^{-1}}$ for Gaia-4b and Gaia-5b, respectively. For Gaia-4b, we elected to remove one observation that had an unusually low SNR (40) due to guiding issues in poor observing conditions, which left 10 observations with a median SNR of 99. All exposures had an exposure time of 900s, except for one of the Gaia-4b observations, which had an exposure time of 245s. Due to the faintness of the targets, we elected not to use the simultaneous laser frequency comb (LFC) calibrator \citep{metcalf2019} and instead performed RV drift corrections following \cite{stefansson2020}. In this procedure, the RV instrumental drifts are estimated from LFC frames taken throughout the observing night and/or right before or after a requested HPF visit on a given target. This methodology has been shown to enable precise wavelength calibration at the $\sim30 \unit{cm\:s^{-1}}$ level, much smaller than the photon-noise dominated RV uncertainty of the observations discussed here.

\subsubsection{NEID}
NEID is a temperature stabilized \citep{stefansson2016,robertson2019} fiber-fed \citep{kanodia2018fiber,kanodia2023} red-optical spectrograph on the WIYN 3.5m Telescope\footnote{The WIYN Observatory is a joint facility of the NSF's National Optical-Infrared Astronomy Research Laboratory, Indiana University, the University of Wisconsin-Madison, Pennsylvania State University, Purdue University and Princeton University.} at Kitt Peak National Observatory in Arizona \citep{schwab2016} covering a wavelength range from 380-930nm at a resolution of $R\sim110,000$ \citep{halverson2016}. The NEID spectra were processed with the NEID Data Reduction Pipeline (DRP)\footnote{\url{https://neid.ipac.caltech.edu/docs/NEID-DRP/}}. The RVs were also extracted from the NEID DRP from the DRP Cross Correlation Functions, where we used the barycentric-corrected radial velocities for re-weighted orders (\texttt{CCFRVMOD}) that we extracted from the NExScI NEID Archive\footnote{\url{https://neid.ipac.caltech.edu/}}.

In addition to characterizing Gaia-4 and Gaia-5, we obtained NEID spectroscopic observations of seven other northern-hemisphere Gaia ASOI candidates around M and K stars. The targets we observed and the results from the observations are further discussed in Section \ref{sec:discussion}. For Gaia-4 and Gaia-5, we obtained six and three observations with a median SNR of 8.7 and 5.2 per extracted 1D pixel at $550 \unit{nm}$, respectively. This resulted in median RV precisions of $8.3\mathrm{m\:s^{-1}}$ and $13\mathrm{m\:s^{-1}}$ for Gaia-4b, and Gaia-5b, respectively. All exposures had an exposure time of 900s, except for one exposure of Gaia-4b, which had an exposure time of 600s. Due to the faintness of the targets, we elected not to use the simultaneous etalon calibrator.

\subsubsection{FIES}
We also employed the FIber-fed Échelle Spectrograph \citep[FIES;][]{Telting+2014} mounted to the $2.56$~m Nordic Optical Telescope, NOT, at the Roque de los Muchachos Observatory on La Palma, Spain. We used FIES in high-resolution mode, which provides a resolving power of $R\approx 67\,000$ with a spectral range of $3700$~\AA{} to $7300$~\AA{}. For stars $<0.7 M_\odot$, we obtained spectra for Gaia-4 (G-ASOI-47), along with G-ASOI-11 and G-ASOI-43. The exposure time used was 2,000s. For Gaia-4 we obtained 5 exposures. We extracted the RVs with the \texttt{SERVAL} code \citep{zechmeister2018} adapted for use on FIES spectra, resulting in a median RV precision of $6\:\mathrm{m\:s^{-1}}$ from the extracted spectra. However, as FIES is not temperature stabilized it is important to account for potential RV drifts during the observations, so we bracketed each observation with a Th-Ar calibration exposure to track the drift. From testing this wavelength correction methodology on Gaia-4 as well as other FIES targets, we adopted a conservative $10\mathrm{m\:s^{-1}}$ wavelength calibration drift uncertainty that we add in quadrature to the RV uncertainties estimated from \texttt{SERVAL}, yielding a total median RV uncertainty of $12.4\mathrm{m\:s^{-1}}$.

\section{Stellar Parameters of Gaia-4 and Gaia-5} \label{sec:stellar}
Table \ref{tab:stellarparam} gives the stellar parameters of Gaia-4 and Gaia-5 drawn from the literature and resulting from our analysis. To obtain spectroscopic constraints on the effective temperature ($T_{\mathrm{eff}}$), metallicity ([Fe/H]), and stellar surface gravity ($\log g$), we used the \texttt{HPF-SpecMatch} code \citep{stefansson2020}. The code implements an emperical spectral matching algorithm that compares a target spectrum to a library of high-SNR observed spectra using a $\chi^2$ metric. From the HPF spectra of Gaia-4, we obtained $T_{\mathrm{eff}}= 4034\pm 77 \unit{K}$, $\mathrm{[Fe/H]}=0.05\pm0.13$, and $\log g = 4.68 \pm 0.05$. These values agree with $T_{\mathrm{eff}}=4000 \unit{K}$ and $\mathrm{[Fe/H]}=0.14$, the values derived by \cite{birky2020} from an analysis of spectra from the Apache Point Observatory Galactic Evolution Experiment \citep[APOGEE;][]{majewski2017}. From the HPF spectra of Gaia-5, we obtained $T_{\mathrm{eff}}= 3447\pm 77 \unit{K}$, $\mathrm{[Fe/H]}=-0.18\pm0.13$, and $\log g = 4.83 \pm 0.05$.

To obtain model-dependent constraints on other stellar parameters, including the stellar mass, radius, and age, we analyzed the two stars using the \texttt{EXOFASTv2} package \citep{eastman2019} using as inputs: a) broadband photometry drawn from the all-sky surveys, b) the Gaia distance from \cite{bailer-jones2018}, and c) the spectroscopic values derived from the HPF data. \texttt{EXOFASTv2} uses the BT-NextGen model grid of theoretical spectra \citep{Allard2012} and the MESA Isochrones and Stellar Tracks \citep[MIST;][]{choi2016,dotter2016} to fit the spectral energy distribution (SED).

We investigated the kinematics of the stars by calculating and interpreting their galactic $U$, $V$, and $W$ velocities with the \texttt{GALPY} \citep{bovy2015} package (Table \ref{tab:stellarparam}). For Gaia-4, the estimated membership probabilities for being part of the thin disk, thick disk, and halo are 99\%, 1\% and $<$1\%, respectively. For Gaia-5, the corresponding probabilities are 98\%, 2\%, and $<$1\%. Thus, both stars are very likely thin-disk members. 

\begin{figure}[t!]
\begin{center}
\includegraphics[width=0.95\columnwidth]{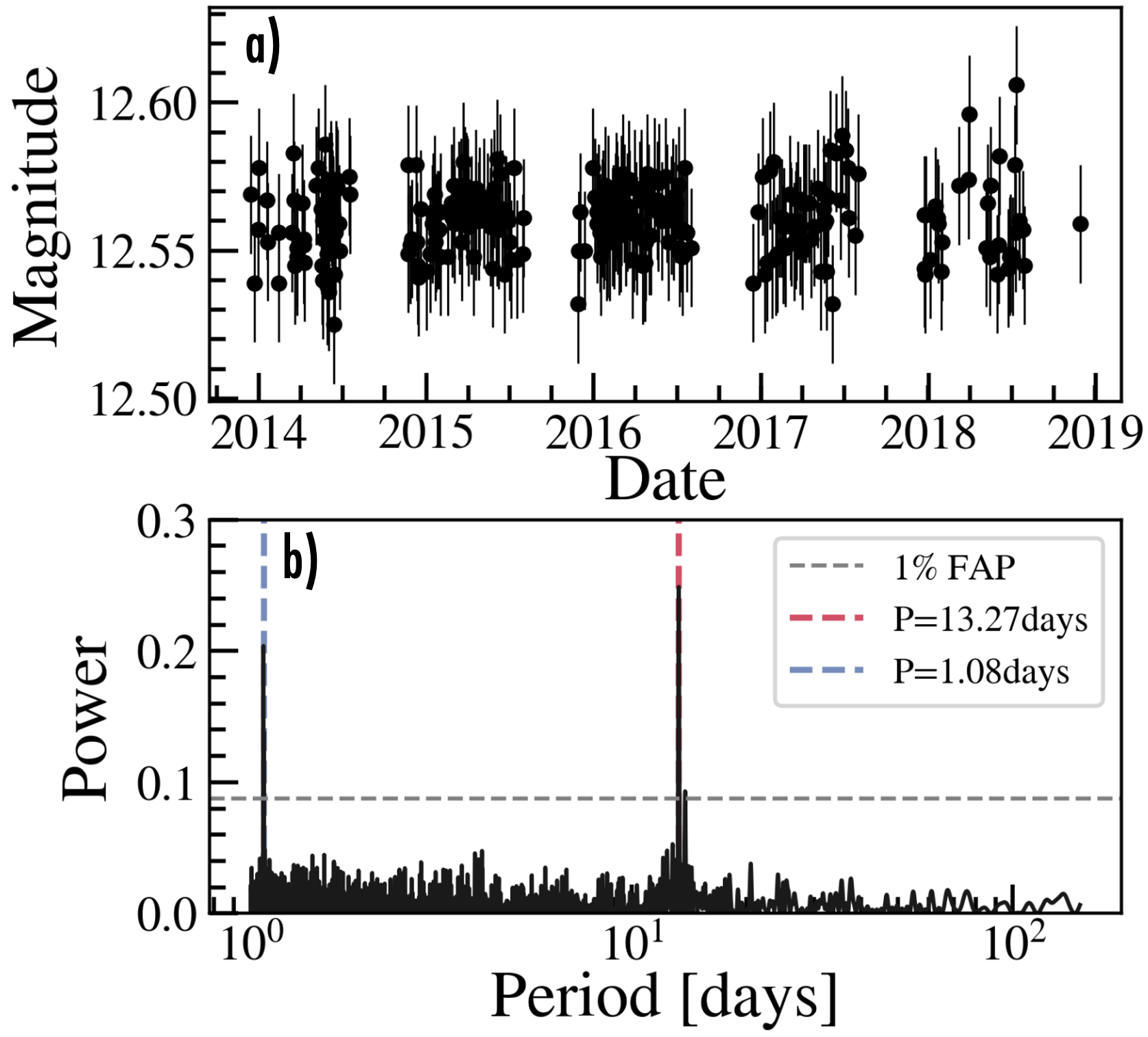}
\vspace{-0.5cm}
\end{center}
\caption{Stellar rotation period for Gaia-4b. a) ASAS-SN photometry as a function of time. b) Generalized Lomb Scargle Periodogram of the photometry in (a) showing a clear peak at 13.3 days (red vertical line) with a false alarm probability (FAP; grey dashed line) $\ll1\%$. The $1 \mathrm{day}$ alias of the 13.3 day period is shown with the blue vertical line. We adopt a stellar rotation period of $13.3 \pm 1.3$ days.}
\label{fig:prot}
\end{figure}

Additionally, we investigated the rotation of the two stars. For Gaia-4, publicly available photometry from the All-Sky Automated Survey for Supernovae \citep[ASAS-SN; ][]{shapee2014,kochanek2017} shows clear modulation with a period of 13.3 days (Figure \ref{fig:prot}). We assigned a 10\% uncertainty to the rotation period to account for potential differential rotation effects. A rotation period of $13.4\pm1.3 \unit{days}$ is also compatible with the observed with the period of 12.97 days reported by \cite{giacobbe2020} (with an amplitude of 5.3mmag and a false alarm probability of 0.2\%) based on ground-based photometric observations from the APACHE survey. For Gaia-5, we investigated available photometric data of the star, including TESS, the Zwicky Transient Facility, and ASAS-SN, but did not find any significant signals indicative of stellar rotation. We note that Gaia-5 has a Kepler Input Catalog ID of 6565450; however, no Kepler data are available for the target.

We used the HPF spectra to measure the projected rotational velocity of both targets, finding $v \sin i = 2.7 \pm 0.8 \unit{km\:s^{-1}}$ and $v \sin i < 2 \unit{km\:s^{-1}}$ for Gaia-4 and Gaia-5, respectively. For Gaia-4, we followed the procedure of \cite{masuda2020} and \cite{stefansson2020k225b} to convert measurements of $v \sin i$ and $P_{\mathrm{rot}}$ into a constraint on the inclination of the stellar rotation axis. The result, $i_\star = 90\pm41^\circ$, is not very constraining.

\begin{deluxetable*}{llcc}
\tablecaption{Summary of stellar parameters used in this work. \label{tab:stellarparam}}
\tabletypesize{\scriptsize}
\tablehead{\colhead{~~~Parameter}                                 &  \colhead{Description}                                                 & \colhead{\gaiaFour}             & \colhead{\gaiaFive}}
\startdata
\multicolumn{4}{l}{\hspace{-0.2cm} Main identifiers:}                                                                                                                                                                       \\
Gaia ASOI ID                                                      &  -                                                                     & Gaia-ASOI-47                    & Gaia-ASOI-15                                 \\
Gaia DR3 Source ID                                                &  -                                                                     & 1457486023639239296             & 2074815898041643520                          \\
TIC                                                               &  -                                                                     & 166669918                       & 42004825                                     \\
2MASS                                                             &  -                                                                     & 2MASS J13580164+3141434         & 2MASS J20080963+4156281                      \\
KIC                                                               &  -                                                                     & -                               & 6565450                                      \\
\multicolumn{4}{l}{\hspace{-0.2cm} Equatorial Coordinates, Proper Motion and Spectral Type:}           \\                                                                     
$\alpha_{\mathrm{J2000}}$                                         &  Right Ascension (RA), epoch J2016                                                  & 13:58:01.53                     & 20:08:09.8                                   \\
$\delta_{\mathrm{J2000}}$                                         &  Declination (Dec), epoch J2016                                                     & +31:41:43.76                    & +41:56:29.56                                 \\ 
$\mu_{\alpha}$                                                    &  Proper motion (RA, \unit{mas\ yr^{-1}})                               & $-75.540 \pm 0.060$             & $102.569 \pm 0.100$                          \\
$\mu_{\delta}$                                                    &  Proper motion (Dec, \unit{mas\ yr^{-1}})                              & $18.069 \pm 0.039$              &  $77.740 \pm 0.095$                          \\
\multicolumn{4}{l}{\hspace{-0.2cm} Equatorial Coordinates, Proper Motion and Spectral Type:}           \\                                                                     
$B$                                                               &  APASS Johnson B mag                                                   & $13.872 \pm 0.069$              & $15.7 \pm 0.031$                             \\
$V$                                                               &  APASS Johnson V mag                                                   & $12.405 \pm 0.069$              & $14.61 \pm 0.2$                              \\
TESS-mag                                                          &  \textit{TESS} magnitude                                               & $11.105 \pm 0.006$              & $12.0628 \pm 0.0073$                         \\ 
\textit{Gaia}-mag                                                 &  \textit{Gaia} magnitude                                               & $11.9187 \pm 0.0004$            & $13.2287 \pm 0.0005$                         \\ 
$J$                                                               &  2MASS $J$ mag                                                         & $9.945 \pm 0.021$               & $10.603 \pm 0.024$                           \\ 
$H$                                                               &  2MASS $H$ mag                                                         & $9.298 \pm 0.014$               & $10.030 \pm 0.017$                           \\ 
$K_S$                                                             &  2MASS $K_S$ mag                                                       & $9.135 \pm 0.019$               & $9.754 \pm 0.018$                            \\ 
WISE1                                                             &  WISE1 mag                                                             & $9.056 \pm 0.024$               & $9.580 \pm 0.022$                            \\ 
WISE2                                                             &  WISE2 mag                                                             & $9.112 \pm 0.020$               & $9.465 \pm 0.020$                            \\ 
WISE3                                                             &  WISE3 mag                                                             & $8.995 \pm 0.028$               & $9.106 \pm 0.033$                            \\ 
WISE4                                                             &  WISE4 mag                                                             & $8.860 \pm 0.353$               & $8.444 \pm 0.285$                            \\ 
\multicolumn{4}{l}{\hspace{-0.2cm} Spectroscopic Parameters$^a$:}           \\                                                                                                
$T_{\mathrm{eff}}$                                                &  Effective temperature in K                                            & $4034 \pm 77$                   & $3447 \pm 77$                                \\
$\mathrm{[Fe/H]}$                                                 &  Metallicity in dex                                                    & $0.05 \pm 0.13$                 & $-0.18 \pm 0.13$                             \\
$\log(g)$                                                         &  Surface gravity in cgs units                                          & $4.68 \pm 77$                   & $4.83 \pm 0.05$                              \\
\multicolumn{4}{l}{\hspace{-0.2cm} Model-Dependent Stellar SED and Isochrone fit Parameters$^b$ (adopted):}           \\                                                             
$M_*$                                                             &  Mass in $M_{\odot}$                                                   & $0.644_{-0.023}^{+0.025}$       & $0.339^{+0.027}_{-0.030}$                   \\
$R_*$                                                             &  Radius in $R_{\odot}$                                                 & $0.624_{-0.015}^{+0.014}$       & $0.345\pm0.013$                             \\
$\rho_*$                                                          &  Density in $\unit{g\:cm^{-3}}$                                        & $3.74^{+0.25}_{-0.24}$          & $11.61^{+0.95}_{-0.88}$                     \\
Age                                                               &  Age in Gyrs                                                           & $6.8_{-4.5}^{+4.6}$             & $8.2^{+4.1}_{-5.0}$                         \\
$L_*$                                                             &  Luminosity in $L_\odot$                                               & $0.1001_{-0.0027}^{+0.0026}$     & $0.01493^{+0.00093}_{-0.00078}$             \\
$A_v$                                                             &  Visual extinction in mag                                              & $0.019^{+0.016}_{-0.013}$       & $0.150^{+0.14}_{-0.097}$                    \\
$d$                                                               &  Distance in pc                                                        & $73.70_{-0.10}^{+0.11}$         & $41.248_{-0.071}^{+0.067}$                  \\
$\pi$                                                             &  Parallax in mas                                                       & $13.628^{+0.021}_{-0.020}$      & $24.208\pm0.036$                            \\
\multicolumn{4}{l}{\hspace{-0.2cm} Other Stellar Parameters:}           \\                                                                                                    
$v \sin i_*$                                                      &  Stellar rotational velocity in \unit{km\ s^{-1}}                      & $<2$                            &  $<2$                                       \\
$P_{\mathrm{rot}}$                                                &  Stellar rotation period in days                                       & $13.3 \pm 1.3$                  &  -                                          \\
$i_\star$                                                         &  Stellar inclination in degrees                                        & $90 \pm 41$                     &  -                                          \\
$RV$                                                              &  Absolute radial velocity in \unit{km\ s^{-1}}                         & $-17.53 \pm 0.23$               & $ -52.70 \pm 0.21$                          \\ 
RUWE                                                              &  Gaia RUWE                                                             & 1.50                            & 3.55                                        \\ 
$U$                                                               &  Galactic $U$ Velocity ($\mathrm{km\:s^{-1}}$)                                          & $-26.21\pm0.05$                 & $-33.78\pm0.06$                             \\
$V$                                                               &  Galactic $V$ Velocity ($\mathrm{km\:s^{-1}}$)                                          & $-15.7\pm0.05$                  & $-45.74\pm0.2$                              \\
$W$                                                               &  Galactic $W$ Velocity ($\mathrm{km\:s^{-1}}$)                                          & $-10.6\pm0.2$                   & $-13.4\pm0.2$                               \\
\enddata
\tablenotetext{}{Magnitudes are from TIC \citep{stassun2018,stassun2019}, Gaia \citep{gaia2018}, and 2MASS/WISE \citep{cutri2014wise}. Distance is from Bailer-Jones \citep{bailer-jones2018}.}
\tablenotetext{a}{Derived using the HPF spectral matching algorithm from \cite{stefansson2020}.}
\tablenotetext{b}{{\tt EXOFASTv2} derived values using MIST isochrones with the Gaia parallax and the HPF-SpecMatch spectroscopic parameters as priors.}
\end{deluxetable*}

\section{Joint RV and Gaia Modeling}
\label{sec:modeling}
To constrain the parameters of the sub-stellar companions we followed a similar path as \cite{winn2022} and \cite{fitzmaurice2023}. We sampled models with three different inputs: i) only the Gaia astrometric solution, ii) only the RVs, and iii) both the Gaia astrometric solution and the RVs. The steps are described below in more detail.

\subsection{`Gaia-only' sampling}
The Gaia two-body solution yields constraints on the following parameters:
\begin{equation}
A, B, F, G, e, P, t_p, \varpi,
\end{equation}
where $A$, $B$, $F$, $G$ are the Thiele-Innes coefficients, $e$ is the eccentricity, $P$ is the orbital period, $t_p$ is the time of periastron, and $\varpi$ is the parallax. To convert between the Thiele-Innes coefficients and the Campbell orbital elements---the argument of periastron ($\omega$), longitude of ascending node ($\Omega$), angular semi-major axis ($a_0$) and the inclination ($i$)---we used the \texttt{nsstools} code. The relevant Campbell elements from the Gaia two-body solutions are listed in Table \ref{tab:paramsmass} and \ref{tab:paramstic}.

To create statistical samples of the posterior probability distribution based on the Gaia covariance matrix, we followed \cite{winn2022} and \cite{fitzmaurice2023} by defining the likelihood function
\begin{equation}
\mathcal{L}_g = \frac{1}{\sqrt{(2\pi)^8|\text{det}\mathcal{C}_{\text{scale}}|}} \text{exp} \left[-\frac{1}{2} (\Theta^{\text{T}} \mathcal{C}_{\text{scale}}^{-1} \Theta ) \right],
\label{eq:likelihoodgaia}
\end{equation}
where $\Theta$ is a vector of deviations between the values reported in the Gaia two-body solution and the sampled value, and $\mathcal{C}_{\text{scale}}$ is the covariance matrix from the Gaia two body solution including a multiplicative factor, $\sigma_{\text{scale}}$, used to optionally scale the original Gaia two-body solution covariance matrix $\mathcal{C}$,
\begin{equation}
\mathcal{C}_{\text{scale}} = \sigma_{\text{scale}}^2 \times \mathcal{C}.
\label{eq:sigma}
\end{equation}
Here, $\sigma_{\text{scale}}$ is a multiplicative factor applied to the tabulated uncertainties. It has the effect of increasing the overall level of uncertainty in the Gaia solution while preserving the relative uncertainties and correlations of the parameters.

\subsection{`RV-only' sampling}
For the RV only-fit, we sampled the following parameters: 
\begin{equation}
P, t_p, e, \omega, K, \gamma,
\end{equation}
where $K$ is the RV semi-amplitude, and $\gamma$ is an RV offset parameter (one for each instrument). We used the likelihood function 
\begin{equation}
\mathcal{L_{\mathrm{v}}} = \prod_{i = 1}^{N} \frac{1}{\sqrt{2\pi(\sigma^2_{v,i})}} \text{exp} \left[-\frac{(v_{i,\mathrm{obs}} - v_{i})^2}{2(\sigma^2_{v,i})} \right],
\label{eq:likelihoodrv}
\end{equation}
where $v_{i,\mathrm{obs}}$ is the $i$-th RV data point and $\sigma_{v,i}$ is the associated uncertainty, and $v_{i}$ is the model evaluated at the same time as the $i$-th data point.

To perform the fit, we first used the differential evolution algorithm implemented in \texttt{PyDE} \citep{pyde} to find the maximum likelihood solution, placing broad uninformative priors on the MCMC parameters. We then use the \texttt{emcee} code \citep{dfm2013}, and initialized 100 walkers around the global maximum solution to perform Markov-Chain Monte Carlo (MCMC) sampling of the posteriors. We ensured that the resulting chains were fully converged and well mixed through visual inspection of the chains, removing initial chains as `burn-in', verifying that the Gelman-Rubin statistic is within 1\% of unity, and following the recommendation given in the \texttt{emcee} documentation to compute the autocorrelation length of the chains and verify that each chain has at least 50 effectively uncorrelated samples. For the final fits, we performed 50,000 MCMC steps, which was sufficient to ensure well-mixed chains.

\subsection{Gaia+RV sampling}
To jointly sample the RVs and the Gaia two-body solution, we followed \cite{winn2022} and \cite{fitzmaurice2023} and use the following parameters as jump parameters in the MCMC sampling:
\begin{equation}
M_\star, m_2, e, \omega, \Omega, \cos i, P, t_p, \varpi, \gamma, \epsilon, \sigma_{\mathrm{scale}},
\end{equation}
where $M_\star$ is the mass of the host star, $m_2$ is the mass of the companion, $\cos i$ is the cosine of the orbital inclination, $\varpi$ is the parallax, $\epsilon$ is parameter denoting the flux ratio between the companion and the host star, and $\sigma_{\mathrm{scale}}$ is the uncertainty scaling factor for the Gaia covariance matrix (see Equation \ref{eq:sigma}). We adopted the joint likelihood function
\begin{equation}
\label{eq:joint_total_likelihood}
\log(\mathcal{L}_{\text{Total}}) = \log(\mathcal{L}_{g}) + \log(\mathcal{L}_{v}),
\end{equation}
where $\log(\mathcal{L}_{g})$ and $\log(\mathcal{L}_{v})$ are given in Equation \ref{eq:likelihoodgaia} and \ref{eq:likelihoodrv}, respectively.

In the sampling of the joint likelihood, the $\sigma_{\text{scale}}$ parameter has the effect of scaling the importance of the two likelihoods relative to each other giving the RV and the astrometric datasets different weights. We considered different fits fixing $\sigma_{\mathrm{scale}}=1$, and fits allowing $\sigma_{\mathrm{scale}}$ to float, to test the agreement of the Gaia two body solution uncertainty estimates with the observed RVs. Due to the good agreement between the RV and the astrometric data for both Gaia-4b and Gaia-5b, we found that those runs resulted in fully consistent parameters. We elected to present the values from the $\sigma_{\mathrm{scale}}=1$ runs, given the simpler model.

To perform the joint sampling, we followed similar steps as the `RV-only' fit discussed above. We used \texttt{PyDE} to find a global maximum likelihood solution with broad uninformative priors on the jump parameters, from which we used \texttt{emcee} to perform MCMC sampling around the global maximum solution. We assessed that the chains were well mixed with similar criteria as discussed in the `RV-only' sampling.

\begin{figure*}[t!]
\begin{center}
\includegraphics[width=0.95\textwidth]{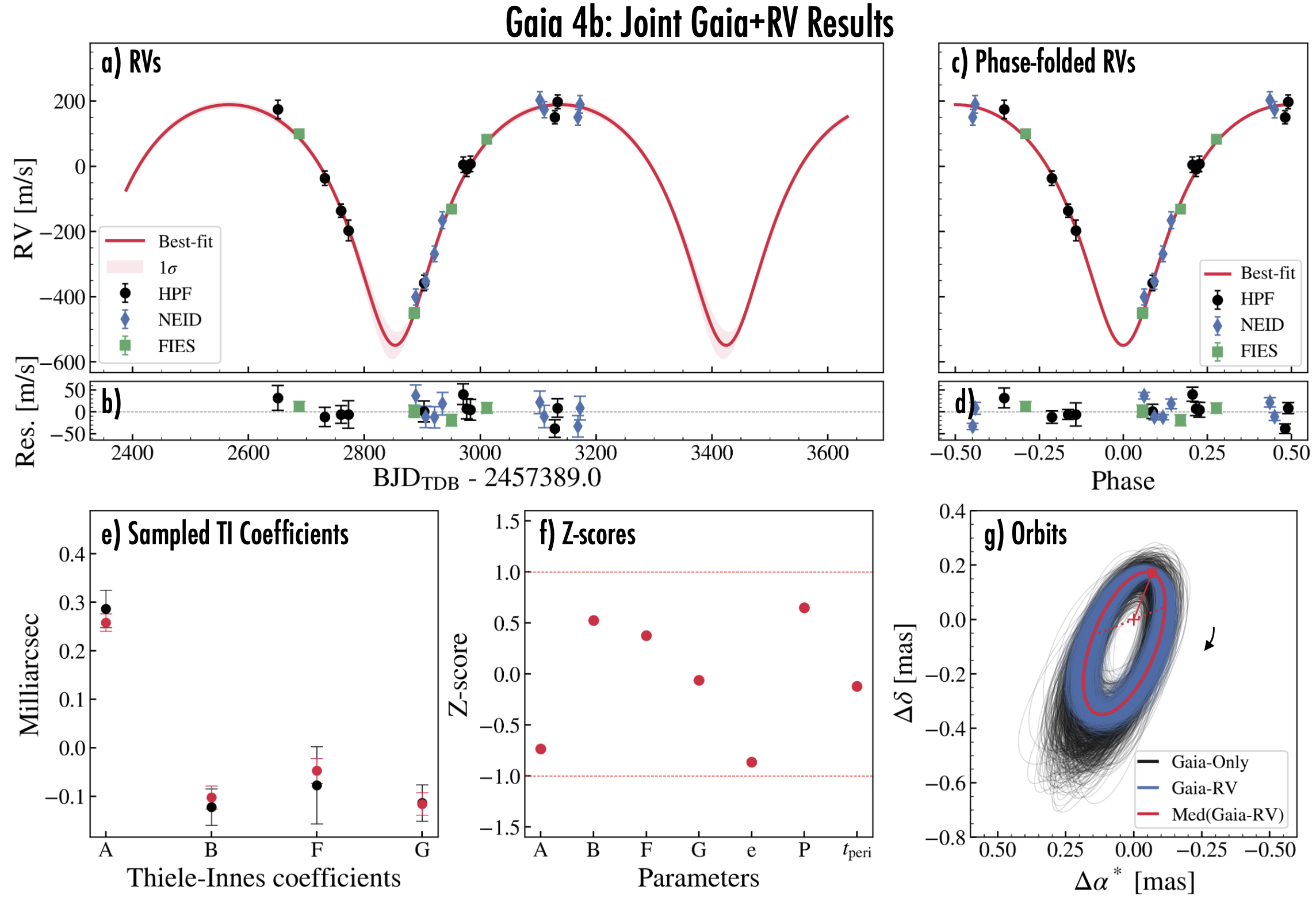}
\vspace{-0.5cm}
\end{center}
\caption{Results from joint sampling of the Gaia two-body solution and RV of Gaia-4b. a) Black data points are from HPF, blue diamond points are from NEID, and green squares are from FIES. The red curve is the best-fit joint Gaia+RV model, and the red shaded region is the 1-$\sigma$ credible interval. b) Residuals. c-d) Same as in a-b) but as a function of orbital phase instead of time. e) Comparison of the Thiele-Innes coefficients in the Gaia-only solution (black) and the joint Gaia+RV solution (red).  f) Z-scores of the Gaia+RV sampled Thiele-Innes coefficients along with $e$, $P$, and $t_p$. In all cases, the Gaia and RV results are in agreement. g) Astrometric orbit of the photocenter, after subtraction of parallax and proper motion. The black and blue curves show 1,000 random draws based on the Gaia-only likelihood and the Gaia+RV likelihood, respectively. The + sign marks the center of mass, and the red square marks the periastron position. The dashed line is the line of nodes, and the arrow indicates the direction of the motion along the orbit. The RVs are available in digital form as `data behind the figure'.}
\label{fig:rvmass}
\end{figure*}

\begin{figure*}[t!]
\begin{center}
\includegraphics[width=0.95\textwidth]{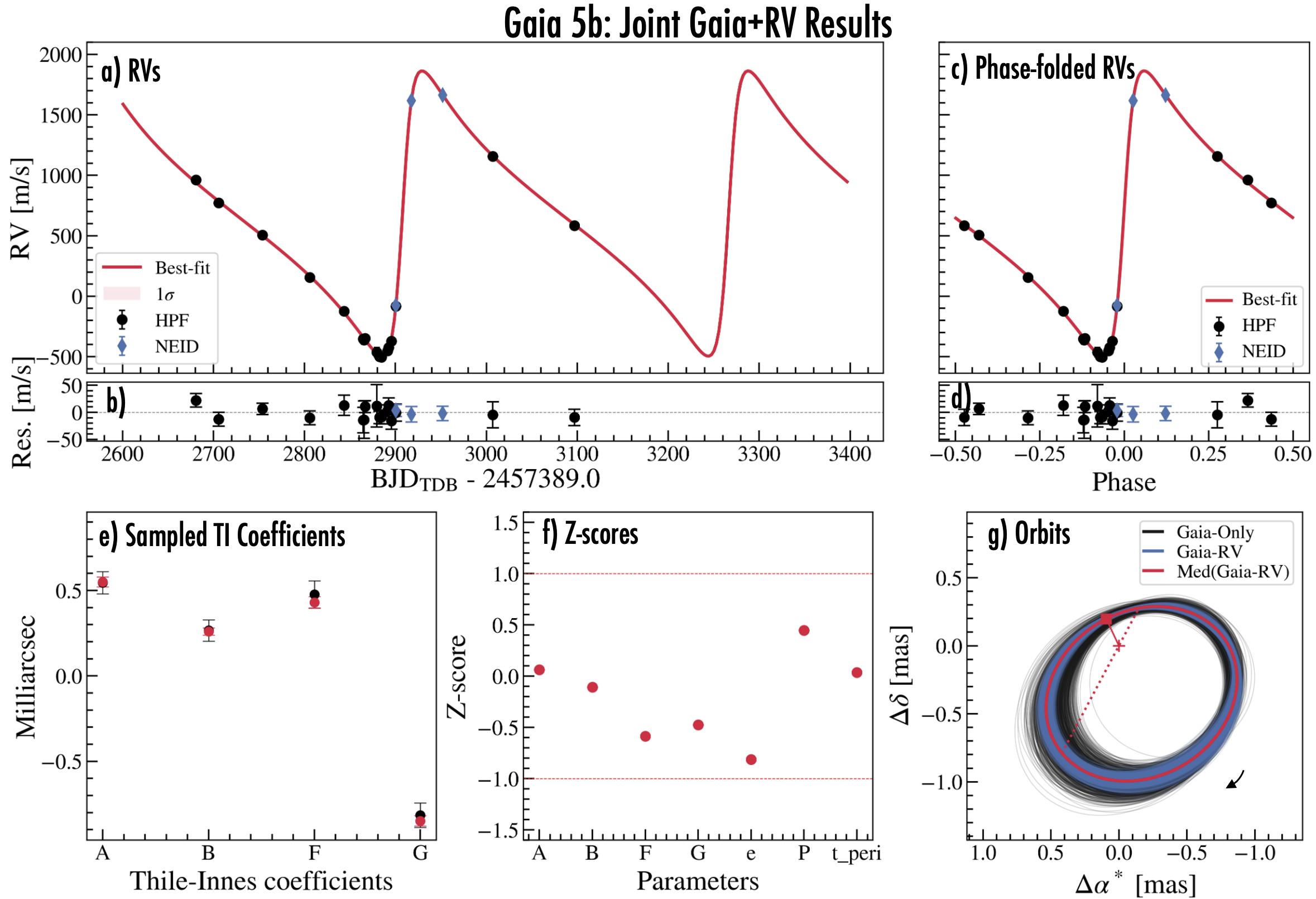}
\vspace{-0.5cm}
\end{center}
\caption{Same as Figure \ref{fig:rvmass}, but for Gaia-5b. The RVs are available as data behind the figure.}
\label{fig:rvtic}
\end{figure*}

\begin{deluxetable*}{llccc}[t!]
\tablecaption{Summary of posteriors for Gaia-4b. Given the good agreement between the Gaia-only solution and the RV-only solution, we adopted the jointly fitted Gaia+RV values as our best estimates of the system parameters. \label{tab:paramsmass}}
\tablehead{\colhead{~~~Parameter} & \colhead{Description}                  & \colhead{Gaia Solution}             & \colhead{RV Only}               & \colhead{Gaia+RV (adopted)}  }
\startdata
$P$                               & Orbital period (days)                  & $564 \pm 11$                        & $513.0_{-16.0}^{+21.0}$         & $571.3_{-1.3}^{+1.4}$                     \\ %
$t_{\mathrm{peri}}$               & Time of Periastron                     & $2457388.6 \pm 20$                  & $2460244.5_{-9.5}^{+10.0}$      & $2457386.1_{-8.4}^{+8.5}$                 \\ %
$e$                               & Eccentricity                           & $0.51 \pm 0.20$                     & $0.263_{-0.033}^{+0.037}$       & $0.338_{-0.023}^{+0.026}$                 \\ %
$\omega$                          & Argument of periastron ($^\circ$)      & $185\pm16$                          & $181.8_{-6.2}^{+6.8}$           & $180.3_{-4.8}^{+5.5}$                     \\ %
$K$                               & Semi-amplitude ($\mathrm{m\:s^{-1}}$)  & -                                   & $355.0_{-14.0}^{+17.0}$         & $368.0_{-18.0}^{+22.0}$                   \\
$\cos i$                          & Cosine of inclination                  & -                                   & -                               & $-0.452_{-0.064}^{+0.070}$                \\
$i$                               & Inclination ($^\circ$)                 & $115.4\pm4.6$                       & -                               & $116.9_{-4.4}^{+4.2}$                     \\ %
$\Omega$                          & Longitude of Ascending Node ($^\circ$) & $159.4\pm5$                         & -                               & $158.6_{-5.1}^{+5.1}$                     \\ %
$\varepsilon$                     & Flux ratio                             & 0                                   & -                               & 0                                         \\ %
$\gamma_{\mathrm{HPF}}$           & RV offset, HPF ($\mathrm{m\:s^{-1}}$)  & -                                   & $-74.0_{-10.0}^{+12.0}$         & $-55.6_{-9.2}^{+10.0}$                    \\
$\gamma_{\mathrm{NEID}}$          & RV offset, NEID ($\mathrm{m\:s^{-1}}$) & -                                   & $-57.0_{-13.0}^{+13.0}$         & $-48.0_{-13.0}^{+13.0}$                   \\
$\gamma_{\mathrm{FIES}}$          & RV offset, FIES ($\mathrm{m\:s^{-1}}$) & -                                   & $42.0_{-13.0}^{+14.0}$          & $67.1_{-10.0}^{+9.9}$                     \\
$\sigma_{\mathrm{HPF}}$           & RV jitter, HPF ($\mathrm{m\:s^{-1}}$)  & -                                   & $14.0_{-12.0}^{+10.0}$          & $17.0_{-11.0}^{+11.0}$                    \\
$\sigma_{\mathrm{NEID}}$          & RV jitter, NEID ($\mathrm{m\:s^{-1}}$) & -                                   & $21.1_{-6.7}^{+10.0}$           & $23.5_{-6.6}^{+10.0}$                     \\
$\sigma_{\mathrm{FIES}}$          & RV jitter, FIES ($\mathrm{m\:s^{-1}}$) & -                                   & $1.3_{-1.0}^{+6.2}$             & $2.5_{-2.3}^{+13.0}$                      \\
$a_0$ (mas)                       & Astrometric amplitude (mas)            & $0.312\pm0.040$                     & -                               & $0.279_{-0.014}^{+0.016}$                 \\ %
$A$                               & Thiele-Innes Coefficient (mas)         & $0.2864 \pm 0.038$                  & -                               & $0.258_{-0.018}^{+0.018}$                 \\ %
$B$                               & Thiele-Innes Coefficient (mas)         & $-0.122 \pm 0.038$                  & -                               & $-0.103_{-0.024}^{+0.024}$                \\ %
$F$                               & Thiele-Innes Coefficient (mas)         & $-0.077 \pm 0.080$                  & -                               & $-0.047_{-0.026}^{+0.025}$                \\ %
$G$                               & Thiele-Innes Coefficient (mas)         & $-0.114 \pm 0.038$                  & -                               & $-0.116_{-0.023}^{+0.023}$                \\ %
$m_2 \sin i$                      & Minimum mass ($M_{\rm J}$)             & -                                   & $10.06_{-0.46}^{+0.51}$         & -                                         \\
$m_2$                             & Companion mass ($M_{\rm J}$)           & $13.2\pm1.9$                        & -                               & $11.8_{-0.66}^{+0.73}$                    \\
\enddata
\end{deluxetable*}

\begin{deluxetable*}{llccc}[t!] 
\tablecaption{Summary of posteriors for Gaia-5b. As noted in the text, the RVs break the discrete degeneracies in $\omega$ and $\Omega$ of the Gaia-only solution. Given the good agreement between the Gaia-only solution and the RV-only solution, we adopted the jointly fitted Gaia+RV values as our best estimates of the system parameters. \label{tab:paramstic}}
\tablehead{\colhead{~~~Parameter} & \colhead{Description}                  & \colhead{Gaia Solution}             & \colhead{RV Only}               & \colhead{Gaia+RV (adopted)}  }
\startdata     
$P$                               & Orbital period (days)                  & $358.0 \pm 1.3$                     & $356.1_{-2.2}^{+2.2}$           & $358.57_{-0.20}^{+0.20}$                  \\ %
t$_{\mathrm{peri}} - 2457389$     & Time of Periastron                     & $39.4 \pm 2.7$                      & $59.0_{-18.0}^{+17.0}$          & $39.5_{-1.6}^{+1.6}$                      \\ %
$e$                               & Eccentricity                           & $0.669 \pm 0.034$                   & $0.639_{-0.0030}^{+0.0030}$     & $0.6412_{-0.0027}^{+0.0027}$              \\ %
$\omega$                          & Argument of periastron ($^\circ$)      & $94.7\pm2.9$                        & $271.71_{-0.65}^{+0.65}$        & $271.72_{-0.59}^{+0.59}$                  \\ %
$K$                               & Semi-amplitude ($\mathrm{m\:s^{-1}}$)  & -                                   & $1177.2_{-9.5}^{+9.4}$          & $1179.8_{-9.0}^{+9.1}$                    \\
$\cos i$                          & Cosine of inclination                  & -                                   & -                               & $-0.638_{-0.014}^{+0.014}$                \\
$i$                               & Inclination ($^\circ$)                 & $129.7\pm2.2$                       & -                               & $129.7_{-1.0}^{+1.0}$                     \\ %
$\Omega$                          & Longitude of Ascending Node ($^\circ$) & $123.3\pm4.9$                       & -                               & $298.0_{-2.5}^{+2.6}$                     \\ %
$\varepsilon$                     & Flux ratio                             & -                                   & -                               & 0                                         \\ %
$\gamma_{\mathrm{HPF}}$           & RV offset, HPF ($\mathrm{m\:s^{-1}}$)  & -                                   & $657.4_{-5.9}^{+5.9}$           & $660.8_{-4.9}^{+4.9}$                     \\
$\gamma_{\mathrm{NEID}}$          & RV offset, NEID ($\mathrm{m\:s^{-1}}$) & -                                   & $-954.0_{-12.0}^{+12.0}$        & $-955.0_{-11.0}^{+11.0}$                  \\
$a_0$ (mas)                       & Astrometric amplitude (mas)            & $0.947\pm0.038$                     & -                               & $0.953_{-0.015}^{+0.016}$                 \\ %
$A$                               & Thiele-Innes Coefficient (mas)         & $0.546 \pm 0.064$                   & -                               & $0.55_{-0.029}^{+0.030}$                  \\ %
$B$                               & Thiele-Innes Coefficient (mas)         & $0.266 \pm 0.063$                   & -                               & $0.26_{-0.022}^{+0.022}$                  \\ %
$F$                               & Thiele-Innes Coefficient (mas)         & $0.477 \pm 0.080$                   & -                               & $0.431_{-0.035}^{+0.035}$                 \\ %
$G$                               & Thiele-Innes Coefficient (mas)         & $-0.816\pm 0.071$                   & -                               & $-0.85_{-0.030}^{+0.031}$                 \\ %
$m_2 \sin i$                      & Minimum mass ($M_{\rm J}$)             & -                                   & $15.35_{-0.89}^{+0.87}$         & -                                         \\
$m_2$                             & Companion mass ($M_{\rm J}$)           & $20.02\pm0.94$                      & -                               & $20.93_{-0.52}^{+0.54}$                   \\
\enddata
\vspace{-0.1cm}
\end{deluxetable*}

\section{Results} 
\label{sec:results}
Tables \ref{tab:paramsmass} and \ref{tab:paramstic} compare the posteriors from the Gaia-only, RV-only, and Gaia+RV jointly fitted models for Gaia-4b, and Gaia-5b, respectively. Figures \ref{fig:corner1}, and \ref{fig:corner2} in the Appendix show corner plots of parameters of interest, showing further that the resulting parameters are in good agreement. For both Gaia-4b and Gaia-5b, the RV-only solutions are in good agreement with the Gaia-only solutions. Figures \ref{fig:expected}a and b in the Appendix compare the astrometry-predicted RV orbit based on the Gaia-only solution with the observed RVs. For both systems, the observed RVs are within the 1-$\sigma$ uncertainty region of the expected RV orbit. This good agreement justified a joint fit of the RVs and the Gaia two-body solution as described in Section \ref{sec:modeling}. The jointly fitted results are shown in Figures \ref{fig:rvmass} and \ref{fig:rvtic} for Gaia-4b and Gaia-5b, respectively. Panel (f) in those figures shows that the Z-scores for the parameters---the number of standard deviations separating the jointly fitted value from the Gaia-only value---have absolute values smaller than unity, another sign of good agreement. 

For the joint fits for Gaia-4 and Gaia-5, we tried letting the flux ratio parameter $\epsilon$ float. To estimate the flux ratio parameter between the primary and the companion in the Gaia bandpass, we performed a 5-degree polynomial fit of mass and Gaia magnitude from the stellar evolution tables for main sequence stars in \cite{pecaut_intrinsic_2013}. We then used this polynomial to evaluate the flux ratio in each MCMC step. Doing so, resulted in posteriors of $\epsilon \sim 10^{-6}$ and $\epsilon \sim 10^{-5}$ for Gaia-4b and Gaia-5b. As these are negligible values, the posterior values for the other parameters were not impacted. In Tables \ref{tab:paramsmass} and \ref{tab:paramstic}, we elected to list the parameters when fixing $\epsilon$ to zero.

For Gaia-4b, the RV jitter values, $\sigma_{\mathrm{HPF}}= 14_{-12}^{+10}\unit{m\:s^{-1}}$, $\sigma_{\mathrm{NEID}}=21.1_{-6.7}^{+10.0} \unit{m\:s^{-1}}$, and $\sigma_{\mathrm{FIES}}=1.3_{-1.0}^{+6.2} \unit{m\:s^{-1}}$ are higher than the median RV uncertainties of $16 \mathrm{m\:s^{-1}}$, $8.3\mathrm{m\:s^{-1}}$, and $12.4\mathrm{m\:s^{-1}}$ for HPF, NEID, and FIES respectively. In other words, the scatter of the RV residuals is higher than the measurement uncertainties. The jitter may be due to stellar activity (e.g., rotationally modulated due to spots moving in and-out-of view), or possibly another companion in the system. The effects of stellar activity would not be surprising, given the clear photometric variation detected in the ASAS-N data (Figure \ref{fig:prot}) with a period of 13.3 days.

As usual for any astrometric orbital solution, the Gaia-only solution cannot distinguish an orbit from its mirror-reflection, with the sky plane as the mirror. \cite{halbwachs2023} noted that for concreteness the Gaia two-body solutions always report values of $\omega$ and $\Omega$ between 0 and $180^\circ$, but the data also permit a solution with both angles advanced by 180$^\circ$. Even a few RVs are enough to break the degeneracy. For Gaia-4b, the RVs imply \resomegagaiafourb, in agreement with the tabulated Gaia-only value of $185 \pm 16^\circ$. For Gaia-5b, the RVs imply \resomegagaiafiveb, which is $180^\circ$ larger than the tabulated Gaia-only value of $\omega = 94.7 \pm 2.9^\circ$. As such, the RVs have broken this degeneracy, suggesting that the true $\omega$ and $\Omega$ values are the ones listed in the Gaia+RV column listed in Table \ref{tab:paramstic}. 

In addition to the consistency checks of the Gaia-only and RV-only solutions, we investigated the likely phase-coverage of the Gaia astrometric observations. Although time-series astrometric data were not part of Gaia DR3, we used the Gaia Observation Scheduling Tool (GOST)\footnote{\url{https://gaia.esac.esa.int/gost/}} to calculate the times when Gaia could have observed either \gaiaFour\ or \gaiaFive. Data were not necessarily obtained at all of the computed times due to telescope downtime or issues with data quality \citep{halbwachs2023}. According to GOST, there were 24 visibility periods for Gaia-4b and 29 visibility periods for Gaia-5b, where a visibility period is defined as a period of observations spanning 4 consecutive days \citep[e.g.,][]{rimoldini2023}. These visibility periods are highlighted as a function of time and overplotted on the astrometric orbit in Figures \ref{fig:astrometry2mass}, and \ref{fig:astrometrytic} in the Appendix. We estimated the individual along-scan astrometric uncertainty to be $0.150 \unit{mas}$ for Gaia-4 and $0.136 \unit{mas}$ for Gaia-5, based on the application of the \texttt{astromet.py} code\footnote{\url{https://github.com/zpenoyre/astromet.py}} which interpolates the median as-obtained Gaia astrometric precision as a function of Gaia magnitude \citep[see Figure 3 of ][]{holl2023}. We additionally assumed that nine individual CCD measurements were obtained in each field-of-view passage \citep{gaia2016}, and that the nine measurements have independent Gaussian uncertainties. The \texttt{gaia\_source} table states that 22 and 28 visibility periods were used for the astrometric solutions for \gaiaFour\ and \gaiaFive, respectively, suggesting that data were not obtained and/or were rejected from only a few visibility periods. 

For both stars, the possible time sampling of the Gaia observations provides good coverage of the inferred astrometric orbit, and the expected astrometric precision seems to be well matched to the range of allowed orbits (see Figure \ref{fig:astrometry2mass}, and \ref{fig:astrometrytic}). Given the good expected phase-coverage of the orbits from Figure \ref{fig:astrometry2mass}, and \ref{fig:astrometrytic}, the derived astrometric solutions seem likely to be trustworthy.

Given the good agreement between the RV-only, and Gaia-only parameters, and the good expected phase-coverage of Gaia of the astrometric orbits, we adopt the jointly fitted Gaia+RV solution values as our best estimates for the system parameters, which are listed in Tables \ref{tab:paramsmass} and \ref{tab:paramstic}.

\begin{figure*}[t!]
\begin{center}
\includegraphics[width=0.8\textwidth]{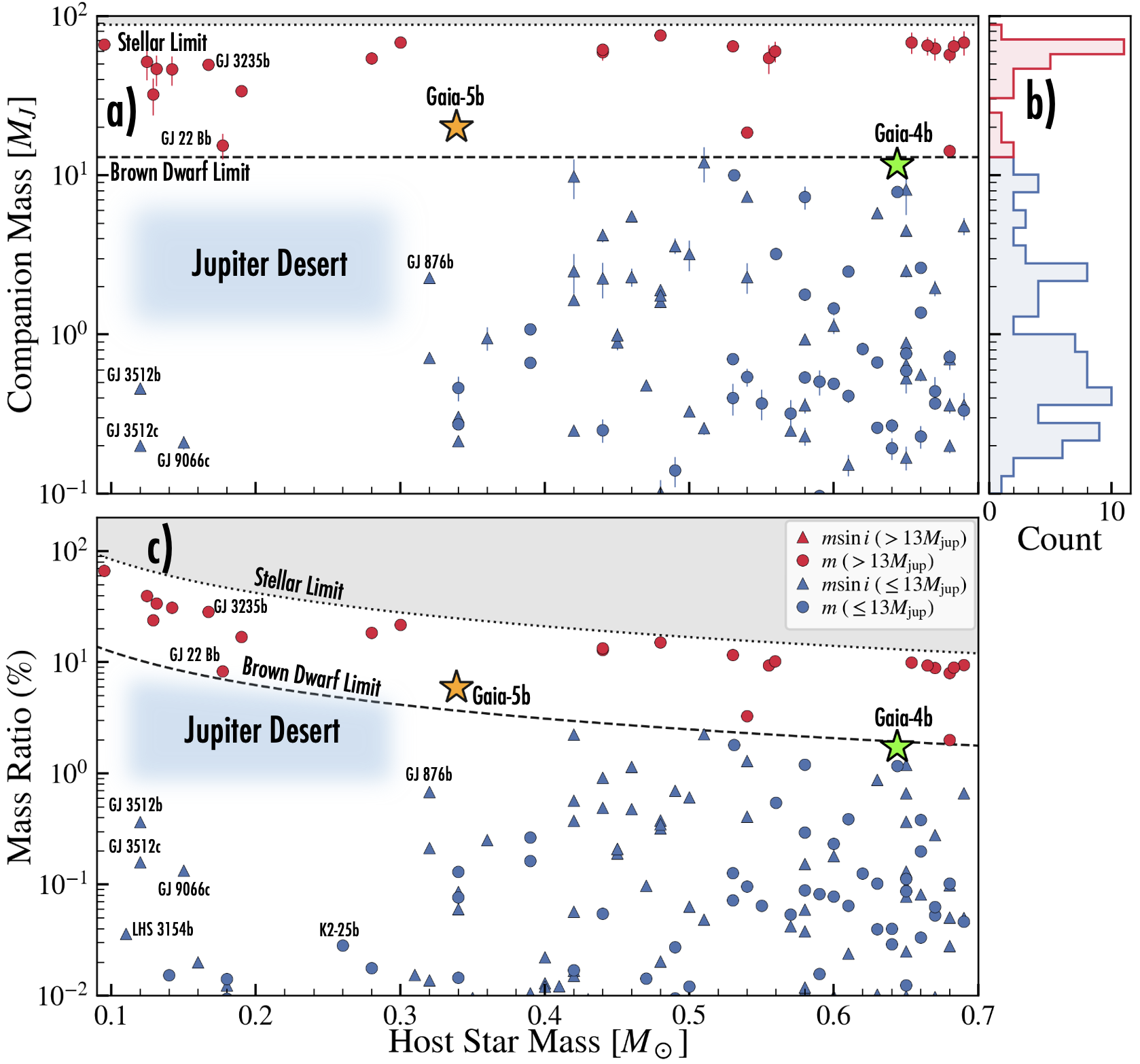}
\vspace{-0.5cm}
\end{center}
\caption{Masses of planets and brown dwarfs as a function of stellar host mass for stars with $<0.7M_\odot$ and orbital periods $<10^4 \unit{days}$. a) Companion mass as a function of host star mass. b) Histogram of the points in panel a). c) Mass ratio as a function of host star mass. The stellar limit ($\sim$80$M_{\mathrm{J}}$) and brown dwarf limit ($13M_{\mathrm{J}}$) are highlighted with the dotted and dashed lines, respectively. Gaia-4b and Gaia-5b are highlighted with the green and orange stars, respectively. Companions with minimum mass measurements ($m \sin i$) are shown with triangles, while true mass measurements ($m$) are shown with the circles. Only companions with better than $3\sigma$ mass measurements are shown. We highlight the lack of $1$ to $\sim$10 $M_\mathrm{J}$ planets around $\leq 0.3 M_\odot$ with the blue `Jupiter desert' region. Planetary data are obtained from the NASA Exoplanet archive \citep{akeson2013}, and brown dwarf data are obtained from Exoplanets.eu, and \cite{stevenson2023}.}
\label{fig:mass}
\end{figure*}

\section{Discussion} \label{sec:discussion}

\subsection{Giant Planets and Brown Dwarfs Around Low-mass Stars}
Gaia-4b and Gaia-5b are substellar objects. We now ask the perennial question, are they `giant planets' or `brown dwarfs'? The distinction of great interest to many researchers is in the formation of such objects \citep{burrows2001,chabrier_giant_2014}. One could use the term `brown dwarf' for a substellar object that forms through instabilities or fragmentation of dense molecular gas either within a disk \citep[e.g.,][]{boss1997,boss2006,boss2023} or via collapse of molecular filaments \citep[e.g.,][]{bate2002,bate2008}, while reserving the term `giant planet' for an object that forms via core-nucleated or pebble-assisted accretion within a disk \citep[e.g.,][]{pollack1996,alibert2005,mordasini2012}. However, it is difficult to determine the formation history of a given object. Further, theoretical investigations into the different formation paradigms have shown that they produce overlapping ranges of companion masses and orbital separations. Gravitational disk instability can form objects as low in mass as Jupiter \citep{boss1997,boss2023}, while core-nucleated accretion can form objects as massive as 10 $M_\mathrm{J}$ under the right conditions \citep{mordasini2012}, blurring the boundary.

An alternative way to differentiate between planets and brown dwarfs is based on whether  deuterium burning occurs within the interior of the object \citep{chabrier2000,burrows2001}. However, deuterium burning cannot be observed directly. Instead, theoretical calculations are used to determine the lower limit on the mass of an object that undergoes deuterium burning. Current calculations lead to an expected lower mass limit for deuterium burning between 11-16$M_{\mathrm{J}}$ depending on the assumed composition \citep{molliere2012}. A somewhat arbitrary and frequently used dividing line is $13 M_{\mathrm{J}}$. From this perspective, Gaia-4b is a giant planet and Gaia-5b is a brown dwarf.

\begin{figure*}[t!]
\begin{center}
\includegraphics[width=0.8\textwidth]{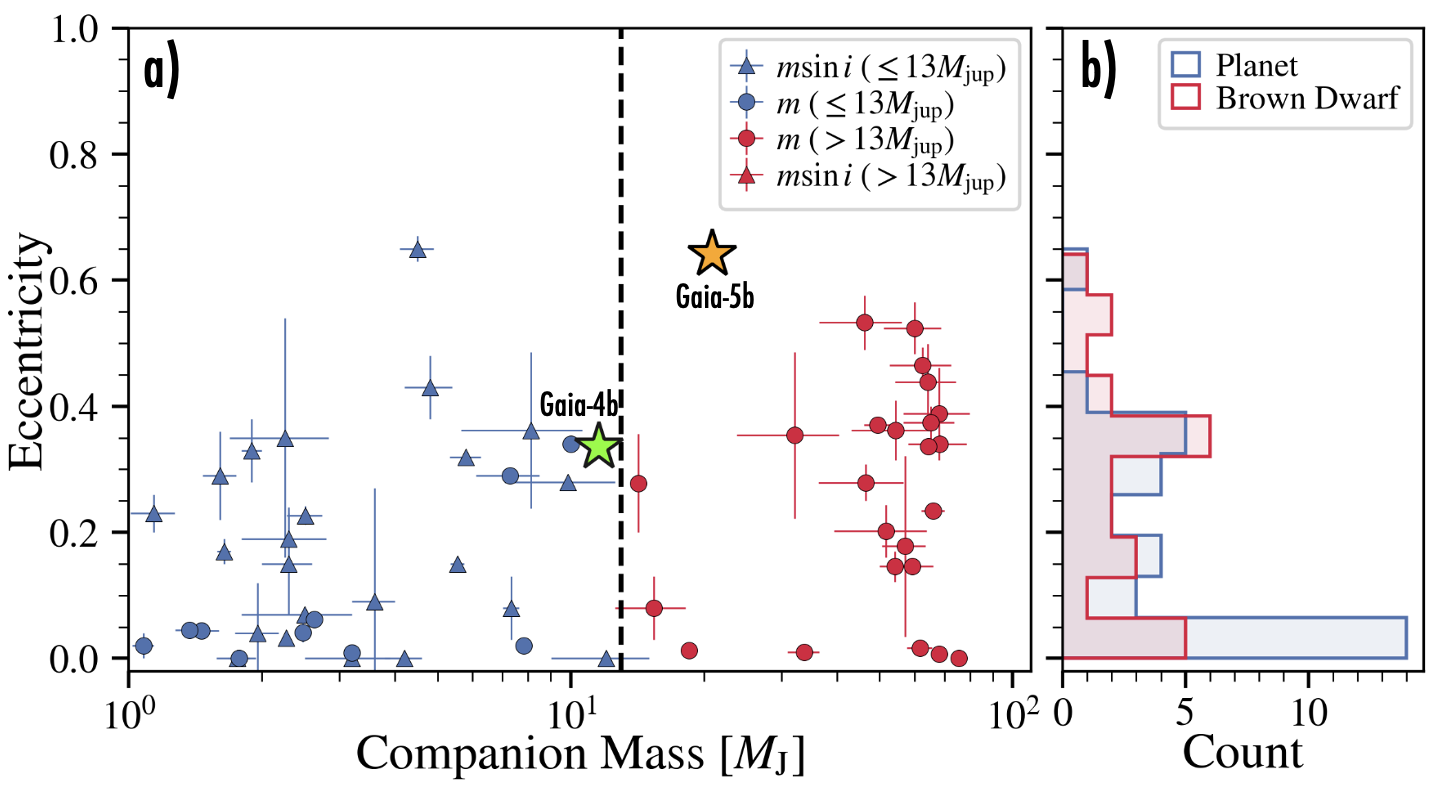}
\vspace{-0.5cm}
\end{center}
\caption{Eccentricities of substellar objects orbiting 0.1--0.7\,$M_\odot$ stars. Only objects for which the statistical significance of the mass measurement is better than 3-$\sigma$ are shown. a) Eccentricity as a function of companion mass. b) Histograms of the orbital eccentricities for companion masses between 1 and 13~$M_{\mathrm{J}}$ (blue) and between 13 and 80~$M_{\mathrm{J}}$ (red). Gaia-4b and Gaia-5b are shown with green and orange stars, respectively. The stellar limit ($\sim$80$M_{\mathrm{J}}$) and brown dwarf limit ($13M_{\mathrm{J}}$) are highlighted with dotted and dashed lines, respectively. Circles represent true mass measurements while triangles are for companions for which only $m \sin i$ is known. The eccentricity distributions of the low-mass and high-mass companions appear different; the low-mass sample includes more circular orbits. Planetary data were obtained from the NASA Exoplanet archive \citep{akeson2013}, and brown dwarf data were obtained from Exoplanets.eu and \cite{stevenson2023}.}
\label{fig:eccentricity}
\end{figure*}

Given the blurred boundaries between massive planets and brown dwarfs, objects in these overlapping windows of massive planets and brown dwarfs present an intriguing opportunity for further study to gain further insights into their formation histories through observations of different observables including orbital distance, eccentricity, and the host star metallicity and mass.

From the perspective of the host star metallicity, core-accretion and gravitational instability predict different observable trends. Core-accretion theory predicts that stars with massive companions should preferentially be metal-rich, since a higher mass of solid material would facilitate the formation of massive companions \citep{fischer_planet-metallicity_2005,santos_observational_2017,osborn2020}. In contrast, gravitational instability does not predict that giant-hosting stars should preferentially be metal-rich \citep[][]{maldonado2017,maldonado2019}. From the analysis of the HPF spectra, Gaia-4 has a metallicity of $\mathrm{[Fe/H]}=0.05 \pm 0.13$ consistent with solar metallicity. This low metallicity for such a massive companion, could therefore favor formation via gravitational instability over core-accretion. For Gaia-5, we measure a sub-solar metallicity of $\mathrm{[Fe/H]}=-0.18 \pm 0.13$, in line with tentative trends of solar and sub-solar metallicities measured for low-mass stars hosting brown dwarfs \citep{maldonado2019}, also pointing towards formation via gravitational instability.

With respect to stellar host star mass, the occurrence of massive planets is known to decrease with decreasing stellar mass \citep{endl2006,johnson_california_2010,maldonado2019,sabotta2021,schlecker2022,gan2023,bryant2023,pass2023}. This has been connected to the fact that less massive stars tend to have less massive protoplanetary disks \citep{pascucci_steeper_2016,ansdell_alma_2016, manara_demographics_2022}, and a lower-mass protoplanetary disk has a smaller inventory of solid materials to nucleate runaway gas accretion. In contrast, brown dwarfs are observed to be rare across all spectral types with an occurrence rate $\lesssim$1\% \citep{sahlmann2011,carmichael2020,barbato2023}, which can be taken as evidence for a different formation pathway for brown dwarfs as compared to giant planets.

Figure \ref{fig:mass} shows the masses of known planets and brown dwarfs orbiting stars with masses between 0.1 and 0.7\,$M_\odot$. A continuum of planets and brown dwarfs has been found around the stars with masses between 0.3 and 0.7\,$M_\odot$ (\autoref{fig:mass}). However, for stars less than $0.3M_\odot$, there is a lack of known Jupiter-mass planets ($1 - 10 M_{\mathrm{J}}$) \citep{pass2023}, highlighted in blue in Figure \ref{fig:mass}. It is unlikely that the gap is due to an observational bias against such systems, as recent RV surveys, such as \cite{sabotta2021} and \cite{pass2023} would have been sensitive to such massive planets around such low-mass stars. Could this `Jupiter desert' reflect a dividing line between different formation channels for brown dwarfs and planets?

Core accretion struggles to form gas giant planets around the lowest mass stars due to the low inventory of dust available along with long formation timescales to form solid cores \citep[e.g.,][]{burn_new_2021,miguel2020,stefansson2023}. To explain the formation of gas giant planets being discovered around low mass stars via photometry from TESS \citep[e.g.,][Hotnisky et al., in prep.]{canas2020,kanodia_toi-5205b_2023,bryant2024} and precise red-optical and near-infrared RV surveys \citep[e.g.,][]{morales2019,quirrenbach2022} necessitates massive protoplanetary disks \citep{delamer_toi-4201_2024}. In such massive disks, formation via gravitational disk instability is an attractive alternative, requiring protostellar disks that undergo rapid gravitational collapse if they are massive and cool enough to satisfy the Toomre criterion \citep{toomre1964,boss2023}. As shown in \cite{boss2023}, the disk-to-star mass ratio required for the protoplanetary disk to undergo collapse increases for lower stellar masses: $\sim$25\% at 0.1 $M_\odot$ compared to $\sim$10\% $\ge$0.3$M_\odot$, suggesting that more massive and rarer disks are required for gravitational collapse to occur around the lowest mass stars. For increasing mass ratios, the likelihood of the cloud fragmenting into separate filaments from which the formation of a star and a massive companion proceeds increases \citep[e.g.,][]{bate2002,bate2008}, representing a favorable mechanism to explain the highest mass ratio brown dwarfs around the lowest mass stars. Surveys such as the ongoing \textit{Searching for Giant Exoplanets around M Stars} survey \citep{kanodia2024}, and MANGOs (\textit{M dwarfs Accompanied by close-iN Giant Orbiters with SPECULOOS}) \citep[e.g.,][]{triaud2023}, along with new astrometric detections from Gaia such as those discussed here will shed further light into the occurrence and rarity of objects within this `Jupiter desert'.

To visualize the eccentricity distribution of massive planets and brown dwarfs comparable to Gaia-4b and Gaia-5b, Figure \ref{fig:eccentricity} shows the measured eccentricities of massive planets ($1M_{\mathrm{J}}<M<13 M_{\mathrm{J}}$) and brown dwarfs ($>13 M_{\mathrm{J}}$) with periods shorter than $10^4 \unit{days}$ around low-mass stars. Gaia-5b has the most eccentric orbit of the brown dwarfs plotted in this figure. Given the orbital separations of both Gaia-4b and Gaia-5b, neither is likely to experience significant tidal circularization. As such, neither system requires a special explanation (such as an additional body) to maintain a high eccentricity. However, how did these objects obtain their eccentricities? Possible scenarios include companion-companion gravitational scattering \citep[e.g.,][]{rasio1996,ford2008,petrovich2016}, secular Kozai-Lidov perturbations with a massive outer companion \citep[e.g.,][]{naoz2016}, or companion-disk interactions during formation \citep{goldreich2003}. From the currently available data, we see no obvious evidence of additional massive companions in the systems. Even so, it still could be a possibility that other companions are present in the systems at distant orbits. To test for potential dynamical influence of a binary star, we checked the Gaia binary star catalog in \cite{elbadry2021} for any co-moving stars, and we found no evidence for such stars. 

From Figure \ref{fig:eccentricity}, both the planet and the brown dwarf sample show similarly distributed tails of eccentric systems, while the planet sample shows a larger number of circular systems. This could hint at a two-component underlying distribution in the planet sample of a close-to an isotropic distribution in high eccentricities combined with a well-aligned component. However, as this is not a homogeneously selected sample, we did not attempt to further characterize the underlying distributions. We expect that in its all sky volume-limited survey, and due to the large number of expected detections of planets and brown dwarfs around low-mass stars, Gaia will enable the creation of such homogeneous samples to shed further light into both the eccentricity distribution between massive planets and brown dwarfs at intermediate orbits. 

\begin{figure*}[t!]
\begin{center}
\includegraphics[width=0.85\textwidth]{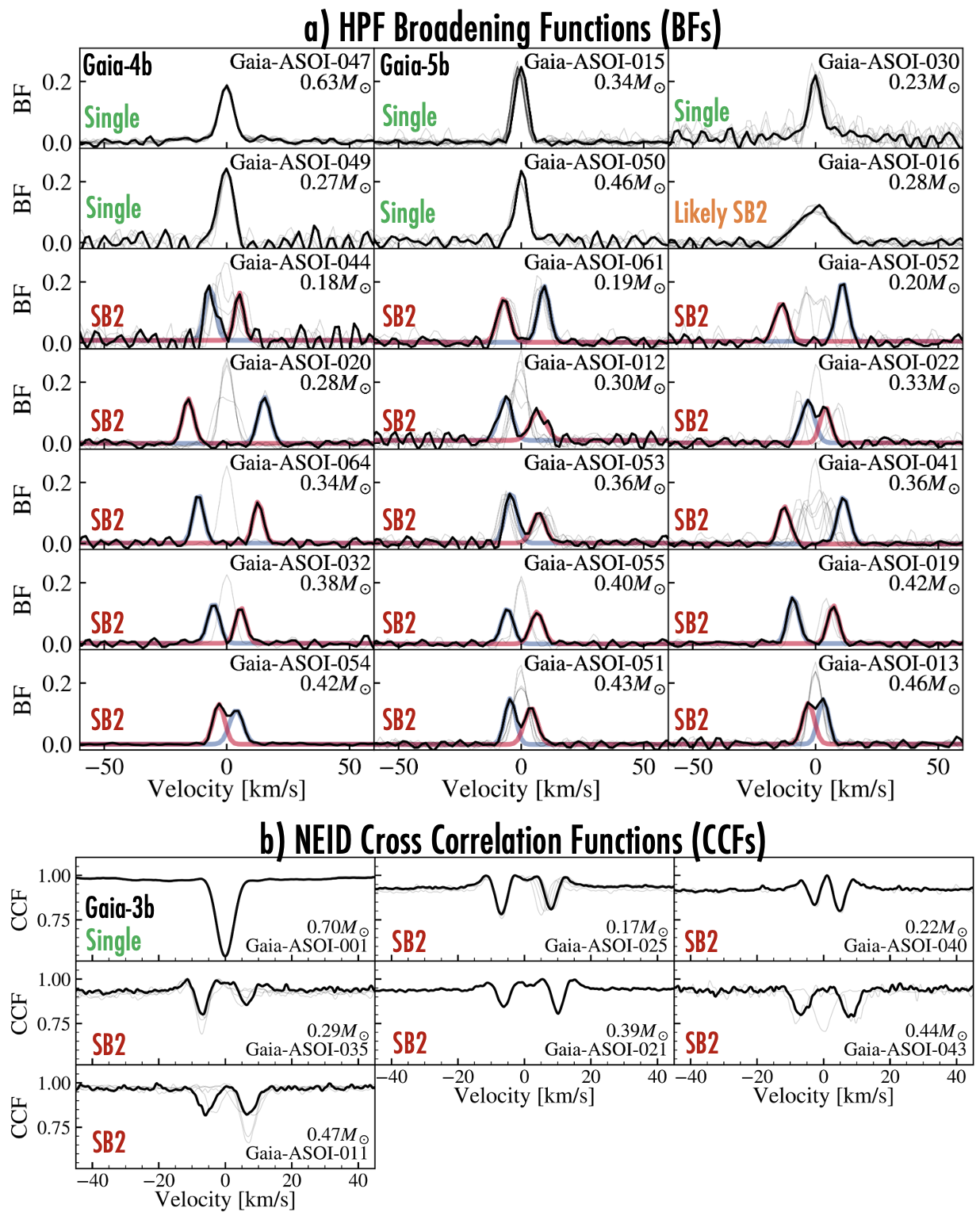}
\vspace{-0.5cm}
\end{center}
\caption{Broadening functions of HPF spectra of the 21 M and K dwarf targets observed as part of the Gaia ASOI exoplanet candidate list. The first 6 show systems thare are compatible with being single lined, although as highlighted in the text Gaia-ASOI-16 could be a SB2 due to the broad spectrum and line width variations. The last 15 show systems that clearly exhibit double lines. If double lines are seen for a system, we pick one such observation where they are clearly seen which we highlight in bold. The faint grey lines show the other epochs we obtained, showing that sometimes the targets are initially seen to be single-lined and later split into two lines. In the cases where two double lines are seen, we additionally show best-fit Gaussians fit to the two peaks, where the blue peak shows the larger peak (larger total integrated area), while the red curve shows the secondary peak. b) Similar to above, but showing the NEID CCFs. Gaia-ASOI-1 is the only single system, whereas the others clearly show double lines. In both a) and b), we first highlight the single-lined systems, and then the SB2 systems, and within those groups the systems are sorted by the presumed host star mass. The host star mass should be taken with caution for the SB2 systems, as it assumes a single star.}
\label{fig:bfs}
\end{figure*}

\subsection{Spectroscopic Insights into Gaia ASOIs} \label{sec:asois}
Including Gaia-4 and Gaia-5, we obtained observations of 28 Gaia-ASOI stars with HPF and NEID orbiting nearby M and K stars. Figure \ref{fig:bfs} shows broadening functions from HPF and CCFs from NEID of those systems. We highlight systems that are consistent with being single stars (i.e., showing only a single set of lines in any of the epochs), as well as systems that show clear evidence of double lines in the spectra. In all cases, when double lines are seen, they have approximately the same height in the BFs or CCFs, confirming that the primary false positive scenario is composed of double-lined nearly equal mass binaries that create small photocenter motions masquerading as planets, similar to the false positive scenario discussed in \cite{marcussen2023}. We note that an observation of a single line is not sufficient to demonstrate that a system is definitely a single star, as it could be that the velocity separation and flux ratios would be such that at the epoch of the observation that the lines would overlap. However, as additional observations are obtained as a function of phase, the possible parameter space for a binary being simultaneously consistent with the astrometric data and the spectroscopic data rapidly shrink. In this work, we present the systems that still show single lines, and we leave the further analysis of them to future work. Additionally, in Figure \ref{fig:bfs} we highlight Gaia-ASOI-016 as a single-lined system that is likely an SB2 system, as although it is formally single lined, the broadening function profiles are very broad and show clear line-width variations suggestive of contaminating light from another star, although additional work would be needed to definitely separate out the two lines.

In Table \ref{tab:targets} in the Appendix, along with Figure \ref{fig:cmd} in Section \ref{sec:obs} we provide an overview of the targets observed and all of the designations (consistent with single, confirmed sub-stellar companion, double lined SB2s) from this work and the literature. Of the 28 systems we observed, 21 are double lined spectroscopic binary systems (SB2s), and six do not show evidence of double lines yet and are thus still compatible with being single stars and are promising candidates for hosting substellar companions consistent with the Gaia astrometric solutions. The last system is Gaia-ASOI-016, which is formally single lined but shows broad time-varying line profiles likely suggestive of an SB2, but we still formally count it in the group of 'single lined' systems. For the promising single-lined systems, additional observations are needed to resolve the full RV curve to independently confirm the RV curve which can then be used to assess the validity of the expected astrometric orbit under the assumption of a single dark companion as we have done for Gaia-4b and Gaia-5b. Such observations are underway. Of the 72 original ASOI candidates, with the confirmation of Gaia-4b and Gaia-5b, in total, 12 are now confirmed to be substellar companions, of which 11 are exoplanets, where nine of the systems were already known RV detections before Gaia. One of the systems is the recently discovered Gaia-3b \citep{holl2023,marcussen2023,winn2022,sozzetti2023} where the ASOI solution is withdrawn, but RV followup suggests a giant planet on an eccentric orbit, and the other two systems are Gaia-4b and Gaia-5b, discussed in this work.

From these observations we see that the false positive rate is between 23/72=32\% and 60/72=83\%, where the former only counts the confirmed SB2s across the full sample, and the latter assumes all non-confirmed exoplanets would be false positives. Restricting our sample of 28 to the 26 systems that did not have any spectroscopic observations before to the best of our knowledge\footnote{The two systems that did have previous observations are G-ASOI-001 (Gaia-3; single lined with likely planet) and G-ASOI-054 (confirmed SB2).}, then the false positive rate is at minimum 20/26 = 77\%, with the remainder being compatible with being single stars, but only 2/26=8\% confirmed substellar companions.

Given the complex selection function steps that went into construction of the ASOI candidate sample, we do not attempt to interpret these numbers as reflective of astrophysical occurrence rates, which would need to wait for robust quantification of biases and sensitivity analysis only possible when the time-series astrometry comes available as part of Gaia DR4. In the meantime, these observations highlight the importance of spectroscopic observations in effectively and efficiently discerning between false positive scenarios. Additional RV observations are ongoing to quantify the agreement with the RV-only solution with the Gaia solution for the other systems that are still compatible with being single.

We note that three systems in the ASOI table, have had their astrometric solution retracted due to a software bug. This includes the polluted white dwarf system WD 0141-675 (Gaia DR3 4698424845771339520) which has been discussed in \cite{rogers2024}, that had a now withdrawn putative substellar candidate with a period of 33.65d. Gaia DR3 1712614124767394816 (Gaia-ASOI-1) is a system discussed by \cite{holl2023,unger2023,winn2022,arenou2023,marcussen2023}. This system was discussed in \cite{winn2022} and \cite{marcussen2023} to show a discrepancy in the RV semiamplitude. Most recently, \cite{sozzetti2023} provided a detailed study in the system, showing that the system has a very high eccentricity of $e=0.95$. They provided simulations suggesting that the Gaia data indicated that the companion would be compatible with being an exoplanet. However, with the formal astrometric solution retracted, the impact on that conclusion has been made unclear. Finally, to the best of our knowledge, Gaia DR3 522135261462534528 (Gaia-ASOI-3) has not been observed spectroscopically.

\section{Conclusion} \label{sec:conclusion}
We announce the discovery of a massive planet and a brown dwarf around nearby low-mass stars using RV follow-up observations of Gaia ASOI exoplanet candidates. 

Gaia-4b is a planet with \resMgaiafour\ orbiting a $0.63M_\odot$ star with an orbital period of \resPgaiafour. Gaia-5b is a brown dwarf with a mass of \resMgaiafive\ orbiting a $0.35 M_\odot$ star with an orbital period of \resPgaiafive. Gaia-4b is the first confirmed planet with a Gaia astrometric solution in agreement with RV observations. This highlights the capability of Gaia to detect planets and substellar companions around nearby stars. For both systems, we show that posteriors derived from an RV-only analysis agrees well with the values expected from the Gaia astrometric two-body solutions. This motivated us to perform a joint sampling of both the RVs and the astrometric solutions, providing a precise characterization of the orbital parameters of the systems. 

Additionally, we discuss spectroscopic observations of 28 Gaia Astrometric Objects of Interests (ASOIs) orbiting nearby M and K stars, where 21 out of the 28 systems are observed to be double lined binaries, and 7 are compatible with being single stars, and two of these 7 are Gaia-4 and Gaia-5. From this, we derive an as-observed false-positive rate of 32-83\% using the full list of 72 Gaia-ASOIs from \cite{holl2023}. Although we do not take this to reflect astrophysical occurrence rates, this highlights the importance of conducting follow-up observations, such as those presented here, to confirm these candidates and rule out false positives masquerading as planetary signals. These detections represent the tip of the iceberg of the planet and brown dwarf yield expected with Gaia in the immediate future enabling key insights into the masses and orbital architectures of numerous massive planets at intermediate orbital periods.

\vspace{1cm}

\textbf{Acknowledgements:} We thank the NEID Queue Observers and WIYN Observing Associates for their skillful execution of our observations. Data presented were obtained by the NEID spectrograph built by Penn State University and operated at the WIYN Observatory by NOIRLab, under the NN-EXPLORE partnership of the National Aeronautics and Space Administration and the National Science Foundation. The NEID archive is operated by the NASA Exoplanet Science Institute at the California Institute of Technology. Based in part on observations at the Kitt Peak National Observatory (Prop. ID 2023B-982758, 2024A-567605, 2024B-726751), managed by the Association of Universities for Research in Astronomy (AURA) under a cooperative agreement with the National Science Foundation. The WIYN Observatory is a joint facility of the NSF's National Optical-Infrared Astronomy Research Laboratory, Indiana University, the University of Wisconsin-Madison, Pennsylvania State University, Purdue University, and Princeton University. The authors are honored to be permitted to conduct astronomical research on Iolkam Du'ag (Kitt Peak), a mountain with particular significance to the Tohono O'odham. Data presented herein were obtained from telescope time allocated to NN-EXPLORE through the scientific partnership of the National Aeronautics and Space Administration, the National Science Foundation, and the National Optical Astronomy Observatory. The research was carried out at the Jet Propulsion Laboratory, California Institute of Technology, under a contract with the National Aeronautics and Space Administration (80NM0018D0004). S.H.A.\ and M.M.\ acknowledge the support from the Danish Council for Independent Research through a grant, No.2032-00230B. CIC acknowledges support by NASA Headquarters through an appointment to the NASA Postdoctoral Program at the Goddard Space Flight Center, administered by ORAU through a contract with NASA. J.I.E.R acknowledges support from ANID BASAL project FB210003 and ANID Doctorado Nacional grant 2021-21212378. S.M., P.R., G.S. and J.L-R. are part of NASA’s CHAMPs team, supported by NASA under Grant No. 80NSSC23K1399 issued through the Interdisciplinary Consortia for Astrobiology Research (ICAR) programme.

This work was partially supported by funding from the Center for Exoplanets and Habitable Worlds. The Center for Exoplanets and Habitable Worlds is supported by the Pennsylvania State University, the Eberly College of Science, and the Pennsylvania Space Grant Consortium. Computations for this research were performed on the Pennsylvania State University’s Institute for Computational \& Data Sciences (ICDS).

These results are based on observations obtained with the Habitable-zone Planet Finder Spectrograph on the HET. We acknowledge support from NSF grants AST 1006676, AST 1126413, AST 1310875, AST 1310885, and the NASA Astrobiology Institute (NNA09DA76A) in our pursuit of precision radial velocities in the NIR. We acknowledge support from the Heising-Simons Foundation via grant 2017-0494. This research was conducted in part under NSF grants AST-2108493, AST-2108512, AST-2108569, and AST-2108801 in support of the HPF Guaranteed Time Observations survey. The Hobby-Eberly Telescope is a joint project of the University of Texas at Austin, the Pennsylvania State University, Ludwig-Maximilians-Universitat Munchen, and Georg-August Universitat Gottingen. The HET is named in honor of its principal benefactors, William P. Hobby and Robert E. Eberly. The HET collaboration acknowledges the support and resources from the Texas Advanced Computing Center. We thank the Resident astronomers and Telescope Operators at the HET for the skillful execution of our observations with HPF.

This work has made use of data from the European Space Agency (ESA) mission Gaia processed by the Gaia Data Processing and Analysis Consortium (DPAC). Funding for the DPAC has been provided by national institutions, in particular the institutions participating in the Gaia Multilateral Agreement.

This research made use of the NASA Exoplanet Archive, which is operated by the California Institute of Technology, under contract with the National Aeronautics and Space Administration under the Exoplanet Exploration Program. This research made use of Astropy, a community-developed core Python package for Astronomy \citep{astropy2013}. This research made use of pystrometry, an open source Python package for astrometry timeseries analysis \citep{sahlmann2019}.

\vspace{5mm}

\facilities{HPF/HET 10m, NEID/WIYN 3.5m, FIES/NOT 2.6m, \textit{Gaia}, \textit{ASAS-N}.} 
\software{\texttt{astropy} \citep{astropy2013},
\texttt{astroquery} \citep{astroquery},
\texttt{barycorrpy} \citep{kanodia2018}, 
\texttt{corner.py} \citep{dfm2016}, 
\texttt{emcee} \citep{dfm2013},
\texttt{pyde} \citep{pyde},
\texttt{radvel} \citep{fulton2018},
\texttt{SERVAL} \citep{zechmeister2018}.}

\newpage

\appendix

\section{Expected RV Orbits from Gaia}
\begin{figure}[H]
\begin{center}
\includegraphics[width=0.95\columnwidth]{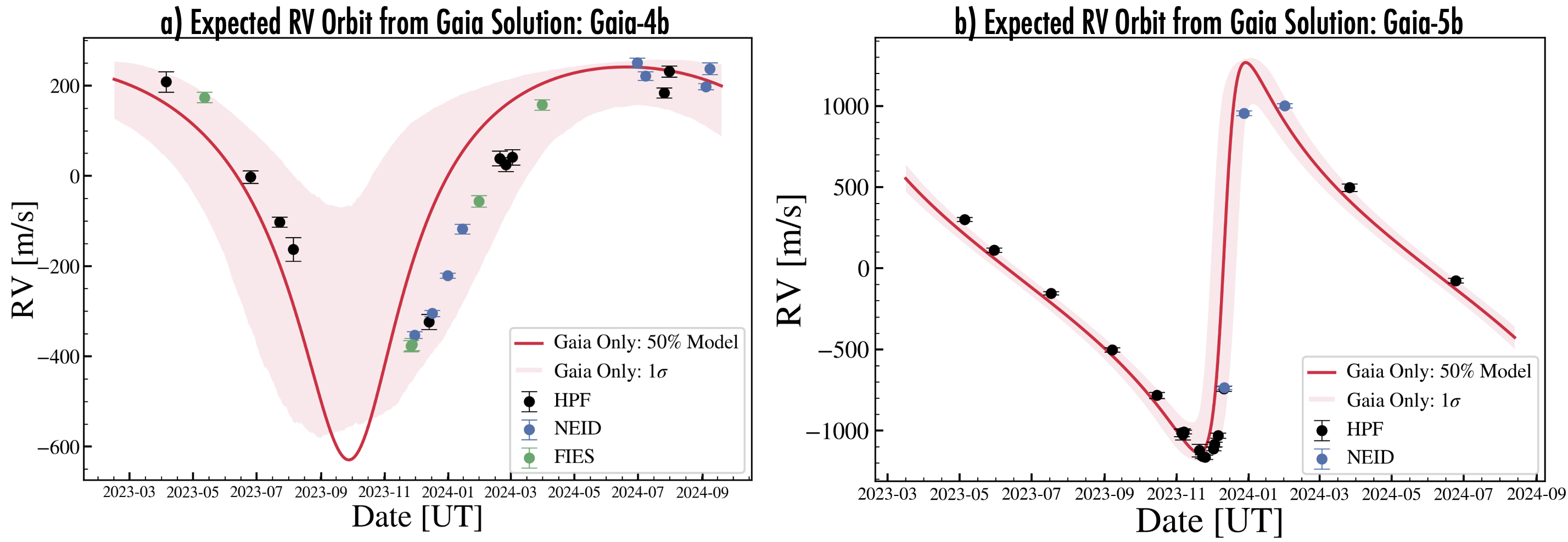}
\vspace{-0.5cm}
\end{center}
\caption{Astrometry-predicted RV orbits for a) Gaia-4b, and b) Gaia-5b compared to the RV observations from HPF (black points), NEID (blue points), and FIES (green points). The red curve shows the expected best model as predicted from the Gaia solution (no fit with the RVs has been performed), and the red shaded region shows the $1\sigma$ region from the expected Gaia orbit. For Gaia-5b, we show the astrometry-predicted RV solution with $\omega + 180^\circ$, due to the $180^\circ$ degeneracy in $\omega$ in the Gaia two-body solution. The RVs agree with the expected RV orbit from the Gaia two body solutions.}
\label{fig:expected}
\end{figure}

Figure \ref{fig:expected} compares the astrometry-predicted RV orbit from the Gaia solution to the RV observations for both Gaia-4b, and Gaia-5b. We see that in both cases, the RVs agree well with the astrometry-predicted RV orbit from the Gaia two body solutions.

\section{Astrometric Time Sampling}
\begin{figure}[H]
\begin{center}
\includegraphics[width=0.8\columnwidth]{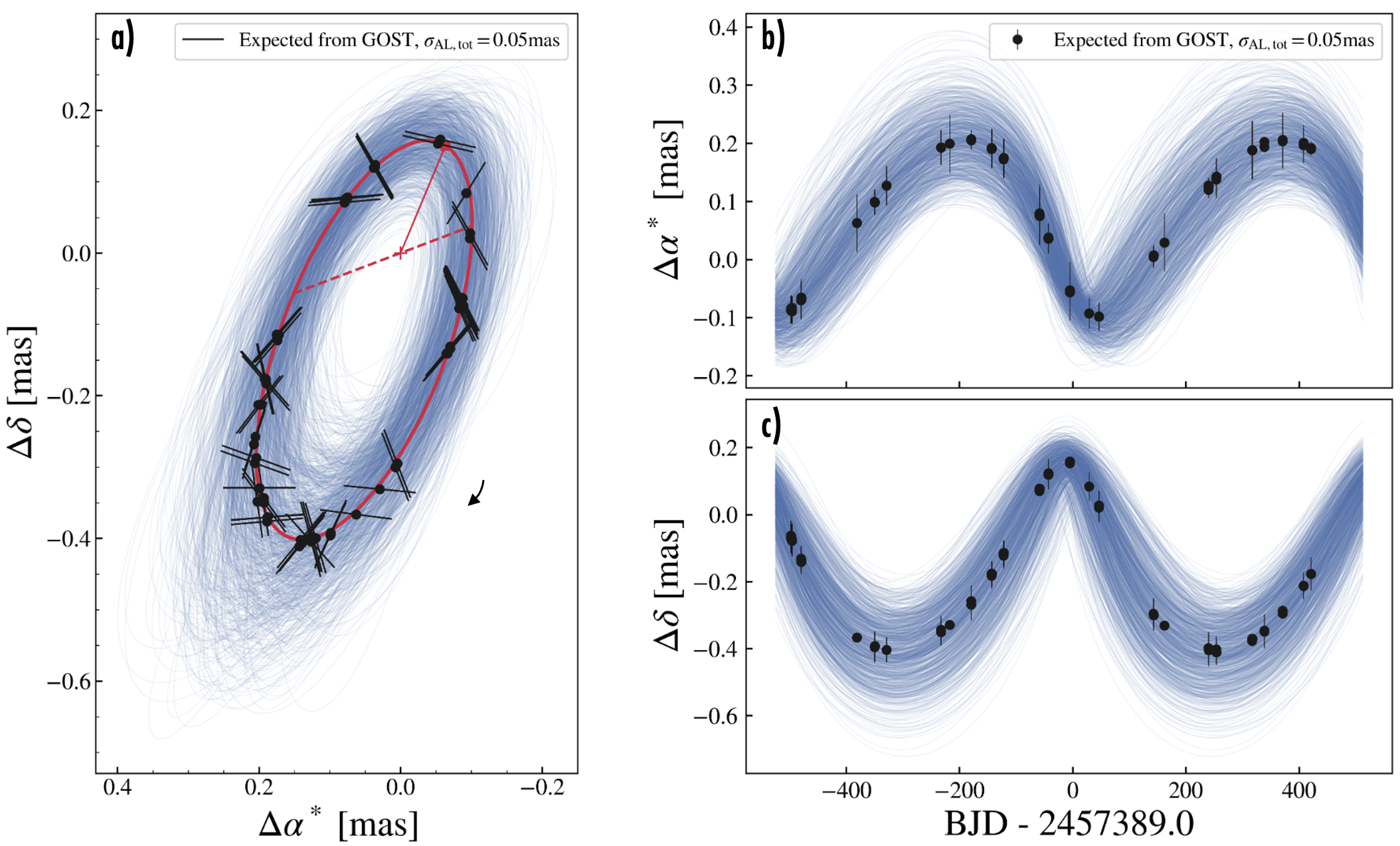}
\vspace{-0.5cm}
\end{center}
\caption{Visualization of the Gaia two-body solution orbit for Gaia-4b. a) Astrometric orbit of the photocenter after a subtraction of parallax and proper motion. The black points show the positions of the photocenter on the sky corresponding to each epoch from the Gaia GOST tool, with uncertainties corresponding to the expected along-scan errors for a star of this brightness scaled by the expected 9 along-scan measurements per epoch. These points are for visualization purposes only. The blue curves show 1,000 random draws of the Gaia orbit. The + sign shows the barycenter, and the red square shows the position of periastron. The dashed line shows the line of nodes, and the arrow indicates the direction of the motion along the orbit. b) Same as a) but showing the right ascension as a function of time. The along-scan uncertainties are scaled in the direction of the right ascension. c) Same as a) and b), but showing the declination as a function of time, with the along-scan uncertainties scaled accordingly.}
\label{fig:astrometry2mass}
\end{figure}

\begin{figure}[H]
\begin{center}
\includegraphics[width=0.8\columnwidth]{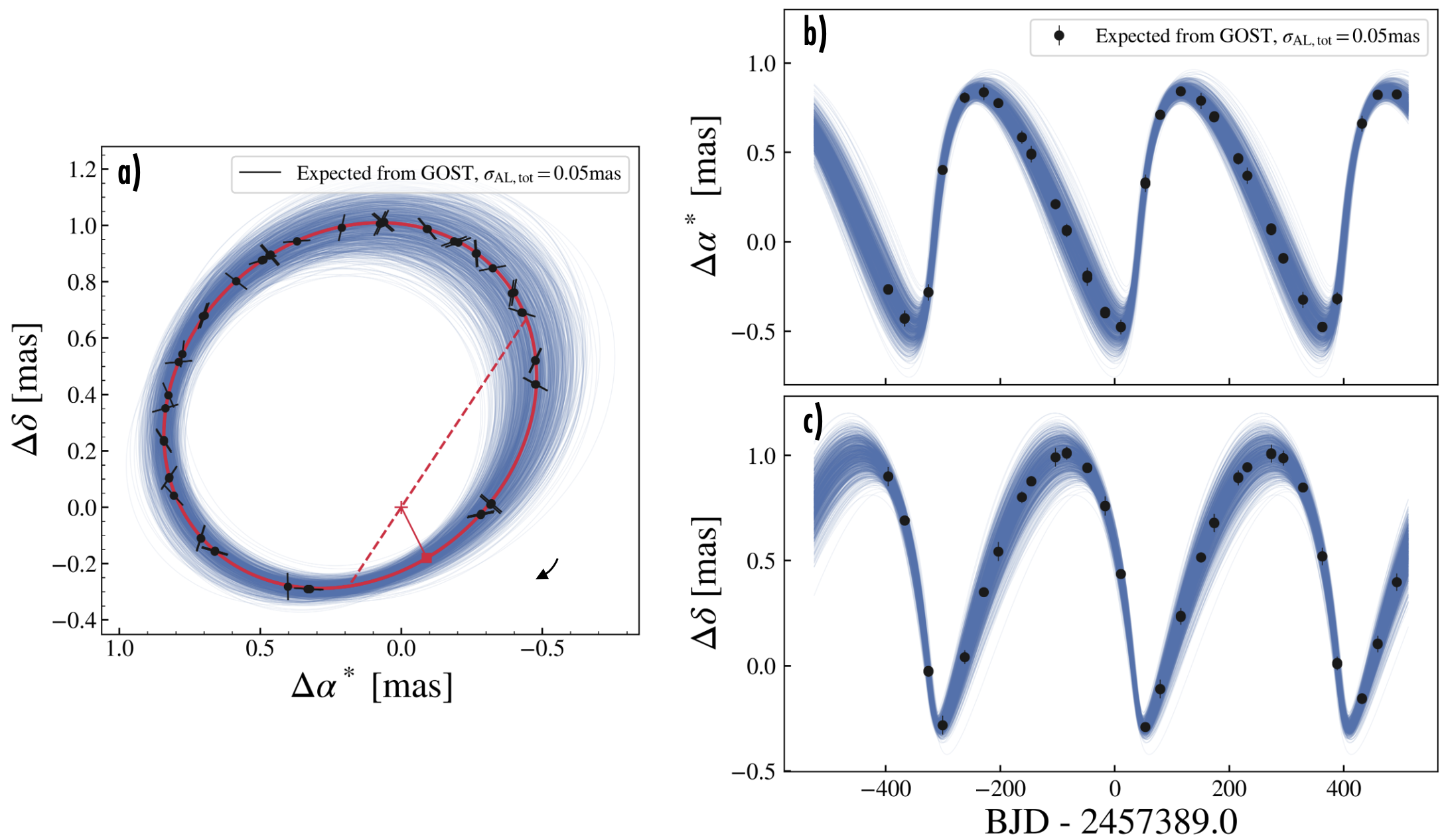}
\end{center}
\vspace{-0.5cm}
\caption{Same as Figure \ref{fig:astrometry2mass} but for Gaia-5b.}
\label{fig:astrometrytic}
\end{figure}

Figures \ref{fig:astrometry2mass} and \ref{fig:astrometrytic} shows the time-sampling of the orbit within the timespan of Gaia DR3 using timestamps from the Gaia scanning law. From this, we see that both systems are well sampled in time. Together with the Gaia scanning law approximately corresponding with the number of astrometric visibility periods used for the two-body solutions, yields further confidence in the solutions.

\section{Corner Plots}

\subsection{Corner plot for Gaia-4b}
\begin{figure}[H]
\begin{center}
\includegraphics[width=0.95\columnwidth]{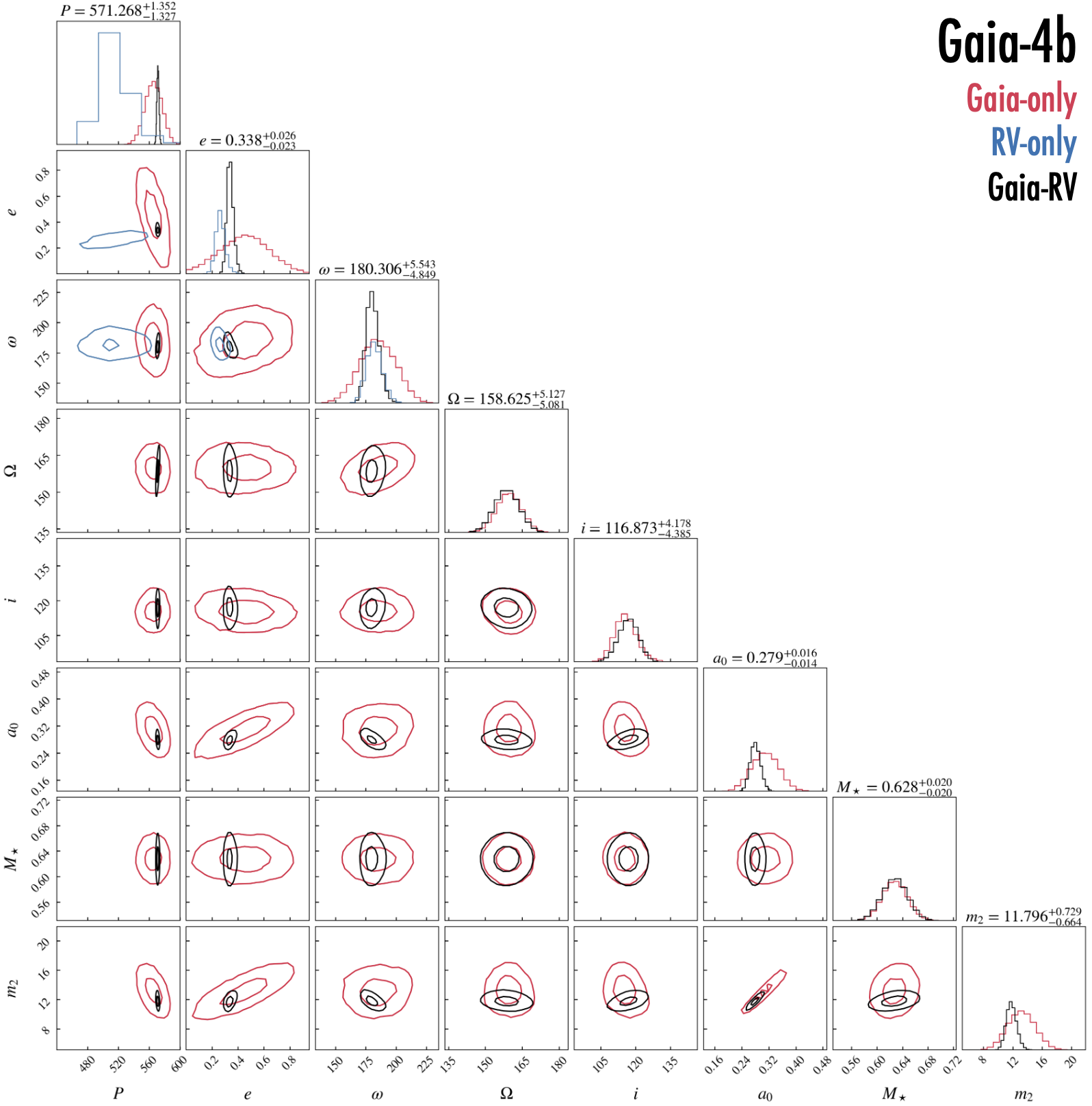}
\vspace{-0.5cm}
\end{center}
\caption{Corner plot for Gaia-4b showing parameters from Gaia-only sampling in red, RV-only sampling in blue, and Gaia+RV sampling in black. The parameters listed at the top of the 1-D histograms are the 16th, 50th and 84th percentile values for the Gaia+RV joint sampling posteriors. The 2D regions show the 1$\sigma$ and 2$\sigma$ contours. We see that all parameters are in good agreement. The orbital period in the RV-only fit shows broad posteriors likely due to only just covering a single orbital period of the system.}
\label{fig:corner1}
\end{figure}

Figure \ref{fig:corner1} shows a corner plot of the posteriors for Gaia-4b. We see that the Gaia-only, RV-only, and the Gaia+RV posteriors are in good agreement.

\subsection{Corner plot for Gaia-5b}
\begin{figure}[H]
\begin{center}
\includegraphics[width=0.95\columnwidth]{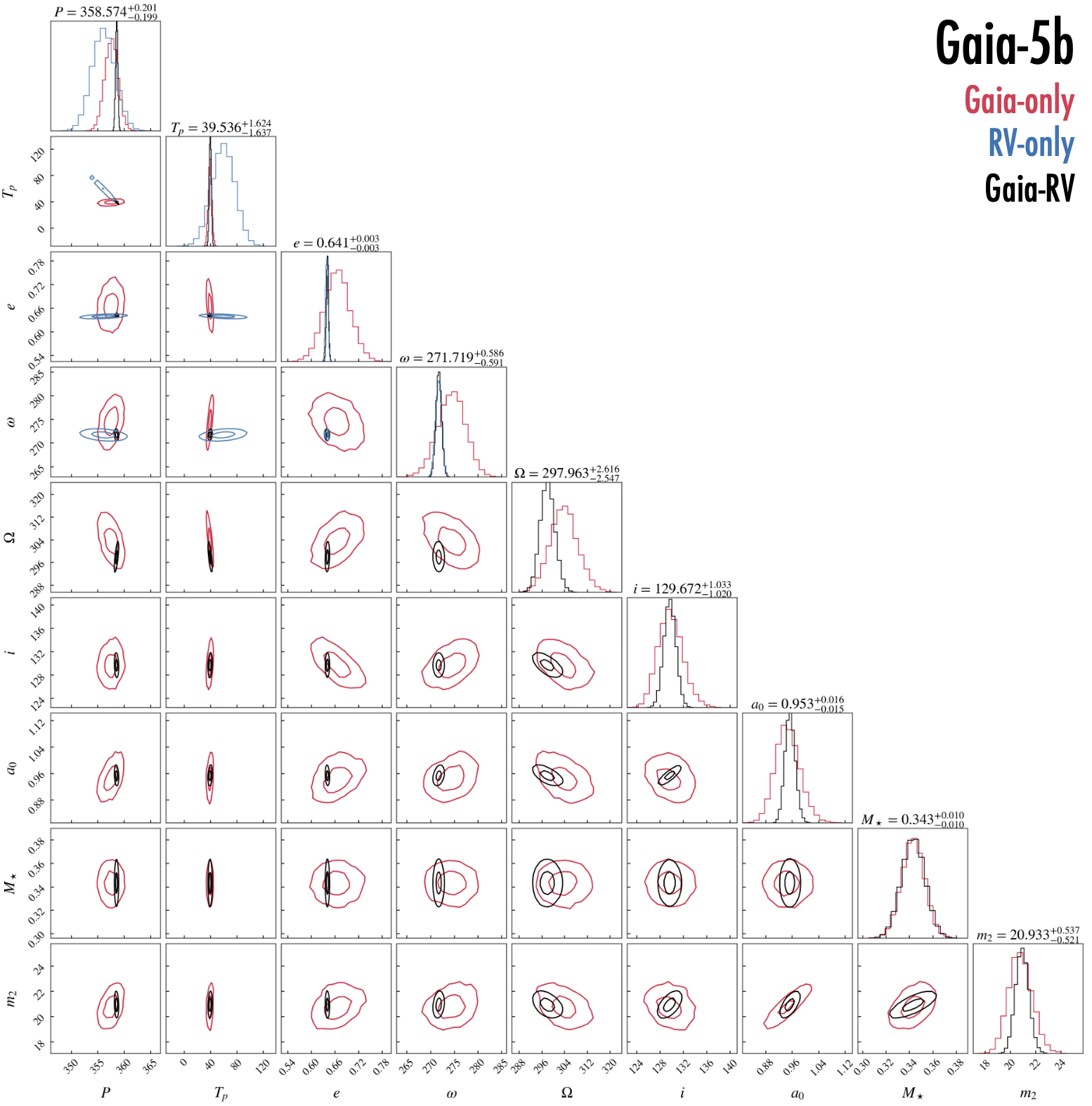}
\vspace{-0.5cm}
\end{center}
\caption{Corner plot for Gaia-5b showing parameters from Gaia-only sampling in red, RV-only sampling in blue, and Gaia+RV sampling in black. For the Gaia-only values, we add $180^\circ$ to both $\omega$ and $\Omega$, to bring them to agreement with the RV-derived values, breaking the known degeneracy in the argument of periastron and longitude of ascending node in the Gaia astrometric solutions. The parameters listed at the top of the 1-D histograms are the 16th, 50th and 84th percentile values for the Gaia+RV joint sampling posteriors. The 2D regions show the 1$\sigma$ and 2$\sigma$ contours. We see that all parameters are in good agreement.}
\label{fig:corner2}
\end{figure}

Figure \ref{fig:corner2} shows a corner plot of the posteriors for Gaia-5b. We see that the Gaia-only, RV-only, and the Gaia+RV posteriors are in good agreement.

\section{Overview of Gaia Exoplanet Candidates}
Table \ref{tab:targets} gives an overview of the Gaia astrometric candidates and designations between exoplanets, single stars, and systems that show evidence of double lines from our spectroscopic follow-up efforts. Figure \ref{fig:bfs} show the broadening functions from our observations showing which systems have double lines, and which are still compatible with single lines.

\startlongtable
\begin{deluxetable*}{rllllllll}
\tablecaption{Gaia Astrometric Exoplanet Candidates from \cite{holl2023}, along with designations from this work and the literature. OrbitalTS and OrbitalTSV stands for \texttt{OrbitalTargetSearch} and \texttt{OrbitalTargetSearchValidated}, respectively. Stellar masses are calculated assuming a single star using the photometric mass relationship from \cite{giovinazzi2022} which is valid for $M\leq1M_\odot$ unless otherwise noted. For the 7 systems with $M > M_\odot$ we quote the mass from the TESS Input Catalog \citep[TIC;][]{stassun2019}. We note that Gaia DR3 4745373133284418816 is listed both as G-ASOI-04 and G-ASOI-68, we only list it as G-ASOI-68. We list Gaia DR3 246890014559489792 as 'single' but note that it has clear line variations, potentially suggestive of a binary star. \label{tab:targets}}
\tablehead{\colhead{~~~Gaia DR3 ID}  &  \colhead{ASOI ID}  &  \colhead{$G$}  &  \colhead{$M_\star$ [$M_\odot$]}  &  \colhead{NSS Solution}  &  \colhead{$P$ [days]}  &   \colhead{$M_2$ [$M_{\mathrm{J}}$]}  & \colhead{Designation}              & Designation Reference  }
\startdata     
1712614124767394816 & G-ASOI-01 & 9.7 & $0.70\pm0.02$  & OTSV$\dagger$ & $297.6\pm2.7$ & $7.1\pm1.1$ & Exoplanet & 1,2,3,4,5,6, This work \\
2884087104955208064 & G-ASOI-02 & 9.0 & $0.84\pm0.03$  & OTSV & $826.5\pm49.9$ & $5.4\pm0.6$ & Single & 1,2,5,6 \\
522135261462534528 & G-ASOI-03 & 6.4 & $1.15\pm0.15$$^{\mathrm{a}}$ & OTS$\dagger$ & $401.1\pm12.0$ & $7.2\pm1.5$ & Unknown &  \\
1878822452815621120 & G-ASOI-05 & 9.8 & $0.88\pm0.04$  & OTS & $1009.3\pm138.5$ & $16.8\pm2.1$ & Unknown &  \\
2047188847334279424 & G-ASOI-06 & 7.3 & $1.10\pm0.06$  & OTS & $450.0\pm5.8$ & $15.3\pm0.9$ & Unknown &  \\
4901802507993393664 & G-ASOI-07 & 9.1 & $0.90\pm0.04$  & OTS & $476.4\pm4.5$ & $13.1\pm0.6$ & SB2 & 1 \\
4698424845771339520 & G-ASOI-08 & 13.7 & $0.58\pm0.02$$^{\mathrm{b}}$ & OTS$\dagger$ & $33.7\pm0.1$ & $2.5\pm0.2$ & Unknown & 7 \\
1610837178107032192 & G-ASOI-09 & 8.2 & $1.17\pm0.16$$^{\mathrm{a}}$ & OTS & $1089.2\pm308.9$ & $11.6\pm14.0$ & Unknown &  \\
4963614887043956096 & G-ASOI-10 & 15.0 & $0.19\pm0.02$  & OTS & $538.0\pm3.0$ & $15.3\pm1.0$ & Unknown &  \\
557717892980808960 & G-ASOI-11 & 12.0 & $0.47\pm0.02$  & Orbital & $522.6\pm8.1$ & $7.6\pm0.6$ & SB2 & This work \\
1862136504889464192 & G-ASOI-12 & 15.2 & $0.30\pm0.01$  & Orbital & $542.4\pm13.5$ & $13.3\pm1.2$ & SB2 & This work \\
2571855077162098944 & G-ASOI-13 & 14.5 & $0.46\pm0.02$  & Orbital & $780.2\pm46.8$ & $13.7\pm1.8$ & SB2 & This work \\
5654515588409756160 & G-ASOI-14 & 16.4 & $0.20\pm0.02$  & Orbital & $774.1\pm60.0$ & $20.7\pm3.3$ & Unknown &  \\
2074815898041643520 & G-ASOI-15 & 13.2 & $0.34\pm0.01$  & Orbital & $357.7\pm1.3$ & $20.0\pm0.9$ & BD & This work \\
246890014559489792 & G-ASOI-16 & 14.4 & $0.28\pm0.02$  & Orbital & $825.2\pm62.4$ & $4.9\pm0.7$ & Single, lines variable & This work \\
4764340705296117120 & G-ASOI-17 & 14.8 & $0.27\pm0.02$  & Orbital & $635.1\pm29.2$ & $10.9\pm0.9$ & Unknown &  \\
5220375041387610880 & G-ASOI-18 & 12.8 & $0.77\pm0.03$  & Orbital & $922.5\pm136.2$ & $19.7\pm4.3$ & Unknown &  \\
1052042828882790016 & G-ASOI-19 & 14.0 & $0.42\pm0.02$  & Orbital & $408.4\pm3.3$ & $19.2\pm0.9$ & SB2 & This work \\
2845310284780420864 & G-ASOI-20 & 13.8 & $0.28\pm0.01$  & Orbital & $417.5\pm6.6$ & $8.5\pm0.6$ & SB2 & This work \\
5085864568417061120 & G-ASOI-21 & 13.7 & $0.38\pm0.01$  & Orbital & $259.6\pm2.3$ & $13.7\pm0.8$ & SB2 & This work \\
423297927866697088 & G-ASOI-22 & 14.8 & $0.33\pm0.01$  & Orbital & $491.9\pm5.5$ & $18.9\pm0.9$ & SB2 & This work \\
6381440834777420928 & G-ASOI-23 & 15.1 & $0.23\pm0.02$  & Orbital & $414.2\pm3.5$ & $18.3\pm1.8$ & Unknown &  \\
6418925831870553472 & G-ASOI-24 & 15.0 & $0.32\pm0.01$  & Orbital & $654.0\pm8.0$ & $13.8\pm0.9$ & Unknown &  \\
4188996885011268608 & G-ASOI-25 & 13.5 & $0.17\pm0.02$  & Orbital & $406.4\pm3.5$ & $6.3\pm0.7$ & SB2 & This work \\
5122670101678217728 & G-ASOI-26 & 8.9 & $1.09\pm0.14$$^{\mathrm{a}}$ & Orbital & $654.0\pm29.8$ & $21.5\pm2.8$ & SB2 & 1 \\
5271515801094390912 & G-ASOI-27 & 12.2 & $0.53\pm0.02$  & Orbital & $622.0\pm5.0$ & $19.9\pm1.0$ & Unknown &  \\
6685861691447769600 & G-ASOI-28 & 14.3 & $0.44\pm0.02$  & Orbital & $645.4\pm31.2$ & $18.5\pm2.5$ & Unknown &  \\
6081071334868194176 & G-ASOI-29 & 12.9 & $0.59\pm0.02$  & Orbital & $423.2\pm9.1$ & $19.3\pm2.1$ & Unknown &  \\
405316961377489792 & G-ASOI-30 & 15.7 & $0.23\pm0.02$  & Orbital & $676.8\pm15.6$ & $20.6\pm3.1$ & Single & This work \\
4842246017566495232 & G-ASOI-31 & 13.0 & $0.31\pm0.01$  & Orbital & $465.4\pm6.5$ & $7.3\pm0.6$ & Unknown &  \\
1879554280883275136 & G-ASOI-32 & 14.3 & $0.38\pm0.02$  & Orbital & $584.5\pm13.0$ & $16.6\pm2.3$ & SB2 & This work \\
6471102606408911360 & G-ASOI-33 & 14.3 & $0.65_{-0.49}^{+0.81}$$^{\mathrm{c}}$ & Orbital & $458.4\pm6.5$ & $10.0\pm0.8$ & Unknown &  \\
5446516751833167744 & G-ASOI-34 & 16.0 & $0.20\pm0.02$  & Orbital & $728.5\pm12.6$ & $13.4\pm1.1$ & Unknown &  \\
2998643469106143104 & G-ASOI-35 & 14.0 & $0.29\pm0.01$  & Orbital & $257.7\pm1.3$ & $17.5\pm0.9$ & SB2 & This work \\
6079316686107743488 & G-ASOI-36 & 14.7 & $0.30\pm0.01$  & Orbital & $557.9\pm9.7$ & $17.6\pm1.3$ & Unknown &  \\
5486916932205092352 & G-ASOI-37 & 12.2 & $0.25\pm0.02$  & Orbital & $253.5\pm0.9$ & $9.0\pm0.5$ & Unknown &  \\
5052449001298518528 & G-ASOI-38 & 13.6 & $0.40\pm0.02$  & Orbital & $336.1\pm3.9$ & $18.0\pm0.9$ & Unknown &  \\
5055723587443420928 & G-ASOI-39 & 14.6 & $0.47\pm0.02$  & Orbital & $546.6\pm8.1$ & $19.4\pm3.8$ & Unknown &  \\
5612039087715504640 & G-ASOI-40 & 13.9 & $0.22\pm0.02$  & Orbital & $592.3\pm3.3$ & $11.9\pm0.6$ & SB2 & This work \\
834357565445682944 & G-ASOI-41 & 13.7 & $0.36\pm0.01$  & Orbital & $480.0\pm6.1$ & $15.0\pm0.9$ & SB2 & This work \\
5490183684330661504 & G-ASOI-42 & 17.7 & $0.13\pm0.01$  & Orbital & $791.0\pm109.4$ & $24.9\pm4.9$ & Unknown &  \\
2277249663873880576 & G-ASOI-43 & 12.9 & $0.44\pm0.02$  & Orbital & $599.4\pm11.8$ & $10.0\pm0.7$ & SB2 & This work \\
2104920835634141696 & G-ASOI-44 & 15.4 & $0.18\pm0.02$  & Orbital & $453.6\pm5.9$ & $11.5\pm0.9$ & SB2 & This work \\
4812716639938468992 & G-ASOI-45 & 14.7 & $0.23\pm0.02$  & Orbital & $519.4\pm4.4$ & $17.9\pm1.1$ & Unknown &  \\
6521749994635476992 & G-ASOI-46 & 13.6 & $0.41\pm0.02$  & Orbital & $337.8\pm4.3$ & $13.3\pm1.0$ & Unknown &  \\
1457486023639239296 & G-ASOI-47 & 11.9 & $0.63\pm0.02$  & Orbital & $564.0\pm11.2$ & $13.2\pm1.7$ & Exoplanet & This work \\
6781298098147816192 & G-ASOI-48 & 14.4 & $0.25\pm0.02$  & Orbital & $602.4\pm11.7$ & $5.4\pm0.5$ & Unknown &  \\
1462767459023424512 & G-ASOI-49 & 15.1 & $0.27\pm0.02$  & Orbital & $781.1\pm33.1$ & $6.2\pm1.1$ & Single & This work \\
2446599193562312320 & G-ASOI-50 & 14.7 & $0.46\pm0.02$  & Orbital & $854.3\pm164.4$ & $22.6\pm4.5$ & Single & This work \\
373892712892466048 & G-ASOI-51 & 13.9 & $0.43\pm0.02$  & Orbital & $477.2\pm5.1$ & $14.9\pm0.9$ & SB2 & This work \\
73648110622521600 & G-ASOI-52 & 15.5 & $0.20\pm0.02$  & Orbital & $491.2\pm7.9$ & $17.8\pm1.2$ & SB2 & This work \\
3937630969071148032 & G-ASOI-53 & 14.7 & $0.36\pm0.01$  & Orbital & $455.4\pm14.0$ & $17.5\pm1.4$ & SB2 & This work \\
2052469973468984192 & G-ASOI-54 & 10.7 & $0.42\pm0.02$  & Orbital & $209.3\pm0.7$ & $16.8\pm0.9$ & SB2 & 1, This work \\
2824801747222539648 & G-ASOI-55 & 14.0 & $0.40\pm0.02$  & Orbital & $560.1\pm6.7$ & $17.3\pm1.3$ & SB2 & This work \\
6694115931396057728 & G-ASOI-56 & 13.4 & $0.30\pm0.01$  & Orbital & $459.3\pm2.4$ & $8.8\pm0.7$ & Unknown &  \\
5618776310850226432 & G-ASOI-57 & 14.7 & $0.39\pm0.02$  & Orbital & $553.5\pm7.8$ & $18.6\pm1.1$ & Unknown &  \\
6677563745912843776 & G-ASOI-58 & 15.5 & $0.29\pm0.01$  & Orbital & $814.9\pm56.8$ & $17.5\pm20.6$ & Unknown &  \\
4702845638429469056 & G-ASOI-59 & 13.9 & $0.39\pm0.02$  & Orbital & $596.6\pm12.5$ & $6.4\pm0.7$ & Unknown &  \\
5375875638010549376 & G-ASOI-60 & 14.4 & $0.33\pm0.01$  & Orbital & $1002.3\pm127.7$ & $12.0\pm2.0$ & Unknown &  \\
3676303512147120512 & G-ASOI-61 & 14.8 & $0.19\pm0.02$  & Orbital & $575.1\pm7.6$ & $15.5\pm1.3$ & SB2 & This work \\
4983571882081864960 & G-ASOI-62 & 13.3 & $0.42\pm0.02$  & Orbital & $227.7\pm1.6$ & $11.3\pm0.7$ & Unknown &  \\
5236626338671861760 & G-ASOI-63 & 13.9 & $0.34\pm0.01$  & Orbital & $505.9\pm9.6$ & $9.9\pm0.7$ & Unknown &  \\
2259968811419624448 & G-ASOI-64 & 13.9 & $0.34\pm0.01$  & Orbital & $171.2\pm0.6$ & $19.2\pm1.1$ & SB2 & This work \\
6421118739093252224 & G-ASOI-65 & 7.8 & $1.00\pm0.13$$^{\mathrm{a}}$ & OTSV & $898.7\pm198.2$ & $8.8\pm1.9$ & Exoplanet & 1,2,4,8 \\
4062446910648807168 & G-ASOI-66 & 9.3 & $0.80\pm0.03$  & OTSV & $615.5\pm12.1$ & $14.0\pm6.2$ & Exoplanet & 1 \\
1594127865540229888 & G-ASOI-67 & 8.3 & $1.03\pm0.13$$^{\mathrm{a}}$ & OTSV & $893.2\pm251.4$ & $6.5\pm2.2$ & Exoplanet & 1,2,4 \\
4745373133284418816 & G-ASOI-68 & 5.3 & $1.17\pm0.16$$^{\mathrm{a}}$ & OTS & $331.7\pm6.2$ & $6.4\pm0.8$ & Exoplanet & 1,2 \\
2367734656180397952 & G-ASOI-69 & 9.2 & $0.77\pm0.03$  & OTSV & $648.9\pm35.6$ & $4.5\pm0.5$ & Exoplanet & 1,2,4 \\
5855730584310531200 & G-ASOI-70 & 7.4 & $1.00\pm0.05$  & OTSV & $882.1\pm33.7$ & $8.6\pm0.7$ & Exoplanet & 1 \\
637329067477530368 & G-ASOI-71 & 7.6 & $1.04\pm0.06$  & OTSV & $850.8\pm112.5$ & $8.2\pm1.0$ & Exoplanet & 1,2,4 \\
4976894960284258048 & G-ASOI-72 & 5.6 & $1.21\pm0.17$$^{\mathrm{a}}$ & OTS & $318.6\pm6.5$ & $7.1\pm1.2$ & Exoplanet & 1 \\
2603090003484152064 & G-ASOI-73 & 8.9 & $0.31\pm0.01$  & OTSV & $61.4\pm0.2$ & $3.1\pm0.4$ & Exoplanet & 1 \\
\enddata
\tablenotetext{}{References are: [1] \cite{marcussen2023}, [2] \cite{unger2023}, [3] \cite{sozzetti2023}, [4] \cite{winn2022}, [5] \cite{holl2023}, [6] \cite{arenou2023}, [7] \cite{rogers2024}, [8] \cite{gan2023rnaas}.}
\tablenotetext{a}{$M_\star$ from TIC.}
\tablenotetext{b}{$M_\star$ from \cite{rogers2024}.}
\tablenotetext{c}{$M_\star$ from the \texttt{binary masses} table from \cite{arenou2023}.}
\tablenotetext{\dagger}{Gaia-ASOI candidates with retracted two-body solutions due to a known software bug, see \url{https://www.cosmos.esa.int/web/gaia/dr3-known-issues}.}
\end{deluxetable*}

\bibliography{references, MyLibrary}{}
\bibliographystyle{aasjournal}

\end{document}